\documentclass{jfm}
\usepackage{array}
\usepackage{epstopdf, epsfig}
\usepackage{comment} 
\usepackage{graphicx}
\usepackage{dcolumn}
\usepackage{bm}
\usepackage{booktabs}
\usepackage[table]{xcolor}
\usepackage{xcolor,colortbl}
\usepackage{amsmath}
\usepackage{float}

\shorttitle{Energy anisotropy over flat terrain}
\shortauthor{I. Stiperski, G. G. Katul, E. Bou-Zeid, M. Calaf}

\title{Stratification, turbulence organization, and pressure-strain effects on surface-layer turbulence anisotropy}

\author{Ivana Stiperski\aff{1}
\corresp{\email{ivana.stiperski@uibk.ac.at}},
G. G. Katul\aff{2},
E. Bou-Zeid\aff{3},
M. Calaf\aff{4}}

\affiliation{\aff{1}Department of Atmospheric and Cryospheric Sciences, University of Innsbruck, Innsbruck 6020, Austria
\aff{2} Department of Civil and Environmental Engineering, Duke University, Durham, North Carolina 27708, USA
\aff{3} Department of Civil and Environmental Engineering, Princeton University, Princeton,
NJ 08544, USA
\aff{4}Department of Mechanical Engineering, University of Utah, Salt Lake City, Utah 84112, USA
}

\begin{document}

\maketitle

\begin{abstract}

At large scales, the Reynolds stress tensor exhibits notable anisotropy, a key feature of all wall-bounded turbulent flows. 
Yet, how the drivers of this anisotropy evolve with shearing and thermal stratification in the atmospheric surface layer (ASL) remains a daunting challenge for theory and models alike. 
Here, the velocity variance budgets are used to explore the evolution of anisotropy in the daytime ASL close to the surface, region known to be problematic for large eddy simulations. A special focus is placed on the importance of slow and rapid pressure-strain correlations and the role of transport on partitioning the turbulent kinetic energy among the velocity components. 
Results obtained from near-surface observations of four datasets over flat and horizontally homogeneous terrain show persistent anisotropy over a wide range of flux Richardson numbers $R_{if}$ and wall-normal distances, and highlight the importance of different processes in three distinct flow regimes, roughly related to dynamic ($|R_{if}|\ll1$), dynamic-convective ($|R_{if}|\sim1$) and convective ($|R_{if}|\gg1$) regimes of the ASL. In particular, close to the surface in the dynamic-convective regime, a drop in wall-normal velocity variance and a substantial increase of spanwise velocity variance are shown to result from the increasing role of pressure transport and rapid distortion, related to turbulence organization. This behaviour is not captured by the classic Rotta closure but requires the inclusion of both rapid pressure-strain and transport terms. In all regimes wall blocking is found to influence turbulence close to the surface, thus requiring the adoption of an anisotropic Rotta model to accommodate its effects.

\end{abstract}


\section{Introduction}
\label{sec:intro}

Anisotropy of the Reynolds stress tensor is one of the fundamental characteristics of most turbulent flows in natural, engineering, and technological applications. At large scales, turbulence rarely attains a kinetic energy state that is equipartitioned between the three velocity components or a state where turbulent stresses disappear. Such an isotropic state would `rob' turbulence of its most important characteristic - efficient momentum transport. In fact, turbulence is almost always anisotropic as a result of the generation of turbulence kinetic energy (TKE) along specific directions. In atmospheric flows, two processes dominate anisotropy production: shear injects energy preferentially in the streamwise direction, while buoyancy acts in the vertical direction, either as a source (unstable stratification) or sink (stable stratification) of TKE. Close to a solid surface, turbulence additionally experiences wall blocking \citep{Pope2000}, limiting the energy in the wall-normal direction \citep{HuntGraham1978}. On the other hand, all turbulent flows experience a pressure-strain interaction that acts to redistribute TKE towards the less energetic velocity components and decorrelate component-wise velocity fluctuations, driving turbulence towards an equipartitioned and stress-free state. This process of `pressure redistribution' remains an area of active research \citep[e.g.,][]{KassinosReynolds2001,Nguyen2013,Ding2018,Homan2024,Yi2025}. The anisotropic state of turbulence for a particular flow, therefore, results from the interplay between the efficiency of the anisotropy generation mechanisms versus its destruction by the pressure-strain interactions. 

Atmospheric boundary layer turbulence is a particularly consequential manifestation of wall-bounded turbulence characterized by a very high Reynolds number and prevalent stratification that has an appreciable impact on turbulence, especially through the evolution of turbulence into large-scale organized structures \citep{Hutchins2012}. In fact, atmospheric turbulence is hardly ever truly neutrally stratified \citep{LiHatchinsMarusic2022}. Stratification influences not only the anisotropy production, but also the pressure-redistribution \citep{Nguyen2013,BouZeid2018}, leading to lingering flow anisotropy even at inertial subrange scales \citep{Praskovsky1993,katul1997energy,Stiperski2021,Chowdhuri2024}. 

Early work by \cite{KaderYaglom1990} suggests that such anisotropic nature of turbulence is of particular interest in the atmospheric surface layer (ASL), where it acts as an additional non-dimensional group in surface-layer scaling \citep{Stiperski2023,Mosso2024,Charrondiere2024,Waterman2025},, as well as in the universal nature of the scalewise relaxation to isotropy \citep{Stiperski2021}. The ability of the degree of anisotropy, quantified by the third invariant of the anisotropy stress tensor \citep{Banerjee2007,CHOI_LUMLEY_2001}, to collapse the scaled variances, gradients, and spectra over a wide range of vastly different terrain complexities --- where basic assumptions of boundary layer theories (horizontal homogeneity, flat terrain, no subsidence) are clearly violated and common scaling approaches fail \citep[e.g.,][]{Stiperski2018,Sfyri2018, Stiperski2019,Finnigan2020} --- means that anisotropy encodes much of the complexity of the flow and surface conditions driving the turbulent flow \citep{Mosso2025,Waterman2025,Chowdhuri2024}. Thus, understanding how specific states of anisotropy in turbulent energy (one-component, two-component, three-component) evolve under common conditions is of particular interest, but as of yet, this evolution is mostly based on plausibility arguments, symmetries, and simplified budgets \citep[cf.][]{Stiperski2018}. A way forward is to combine the budget equations that describe the evolution of the components of the Reynolds stress tensor with multi-level measurements collected over flat and horizontally homogeneous terrain, with the goal of unravelling the anisotropy drivers using revisions to conventional turbulence modelling perspectives. 
 
To test how the common Reynolds stress budget models used in higher-order closures capture observed flow anisotropy, we leverage turbulence measurements from a range of observational campaigns. As a first approximation, we test how a reduced set of budget equations captures the observed energy anisotropy, i.e., how energy is distributed between the different velocity variances. This reduced model assumes a balance between shear and buoyancy production and dissipation, and models the return to isotropy using a linear Rotta scheme. This model provides  expressions for normalized velocity variances as functions of the Richardson number, highlighting the change of anisotropy as stratification becomes progressively more dominant. These expressions then serve as a reference to an expanded version of the model that includes transport terms,, as well as the pressure-strain parametrization by adding the so-called rapid isotropization of the production terms. Finally, the origins of the transport and rapid terms are explored.  

The paper is organized as follows: In Sect. \ref{sec:models} the Reynolds stress budgets, simplifications, and closure assumptions are presented, in Sect. \ref{sec:data} the datasets and turbulence data processing used are explained. Section \ref{sec:drivers} presents the results, further discussed in Sect. \ref{sec:discussion}, with conclusions presented in Sect. \ref{sec:conclusions}.  

\section{Modelling the Reynolds stresses}
\label{sec:models}

\subsection{Background and definitions}
\label{sec:reynoldsstressbudget}
The conservation equations for the Reynolds stresses $\overline{u_i' u_j'}$ for an incompressible flow subject to the Boussinesq approximation, a linearized equation of state for air (ideal gas law), and hydrostatic equilibrium for the Boussinesq background state are given by \citep{launder1975progress,Stull88,Pope2000},
\begin{eqnarray}
  \frac{\partial\overline{u_i'u_j'}}{\partial t}+\overline{U}_k\frac{\partial\overline{u_i'u_j'}}{\partial x_k} = && \underbrace{-\overline{u_{i}'u_{k}'}\frac{\partial\overline{U}_j}{\partial
    x_k}-\overline{u_{j}'u_{k}'}\frac{\partial\overline{U}_i}{\partial
      x_k}}_{2S_m}
      \underbrace{+\frac{g}{\overline{\theta}_{v0}}\left[\overline{u_i'\theta_v'}\delta_{j3} + \overline{u_j'\theta_v'}\delta_{i3}\right]}_{2B} 
      \label{eq:Reynolds} \\
  && 
\underbrace{+f_c\left[\overline{u_i'u_k'}\epsilon_{jk3} + \overline{u_j'u_k'}\epsilon_{ik3}\right]}_{Co} \underbrace{-\frac{\partial\overline{u_i'u_j'u_k'}}{\partial x_k}}_{2T} \nonumber\\
  &&
\underbrace{\underbrace{-\frac{1}{\rho_0}\left[\frac{\partial\overline{u_i'p'}}{\partial x_j} + \frac{\partial\overline{u_j'p'}}{\partial x_i}\right]}_{\Pi_{ij}^t}
  + \underbrace{\frac{1}{\rho_0} \left[\overline{p' \left( \frac{\partial u_i'}{\partial x_j} + \frac{\partial u_j'}{\partial x_i} \right) }\right]}_{\Pi_{ij}^s}}_{\Pi_{ij}}  -2\mathrm{\varepsilon_{u_i u_j}},  
  \nonumber
\end{eqnarray}
where $t$ is time, $\rho_o$ is the Boussinesq reference air density, $U_i=\overline{U_i}+u_i'$ are the instantaneous velocity components along the $x_i$ direction, where $x_1$, $x_2$, and $x_3$ are the streamwise (along mean wind direction), spanwise, and wall-normal directions, respectively, and their corresponding velocity components are $u$, $v$, and $w$. Overline indicates Reynolds-averaged flow variables, primed quantities are fluctuations from their respective Reynolds-averaged state, $p'$ are the corresponding pressure fluctuations (the hydrostatic background pressure is already removed along with the gravity term), $\theta_v'$ are the virtual potential temperature fluctuations (with their reference state ${\overline{\theta}_{v0}}$ corresponding to $\rho_0$), and $\mathrm{\varepsilon_{u_i u_j}}$ is the mean turbulent stress destruction rate due to the action of fluid viscosity $\nu$. The third term is a Coriolis redistribution term, which is not identically zero for the individual variances, but sums to zero for the TKE. In the budget equations, closure models for $\mathrm{\varepsilon_{u_i u_j}}$, the turbulent transport ($T$) terms 
, and the pressure-covariance term $\Pi_{ij}$ (which can be decomposed into the pressure transport $\Pi_{ij}^t$ and pressure-strain $\Pi_{ij}^s$) are necessary. The most difficult and least understood among all these unclosed terms is $\Pi_{ij}$, which is commensurate in magnitude with the turbulence generation terms (the shear production, first term on the right-hand side of the equation, and buoyancy production/destruction, the second term). The reason for the difficulty in modelling $\Pi_{ij}$ is the fact that pressure perturbations $p'$ satisfy the Poisson equation given by \citep{hanjalic1972reynolds,launder1975progress}
\begin{equation}
\label{eq:p_Poisson}
\frac{1}{\rho_o} \nabla^2 p'= \underbrace{-2 \frac{\partial \overline{U_i}}{\partial x_j} \frac{\partial u_j'}{\partial x_i}}_{rapid~term} \underbrace{-\frac{\partial^2}{\partial x_i \partial x_j} \left(u_i' u_j' - \overline{u_i' u_j'}\right)}_{slow~term}+\underbrace{\frac{g}{\overline{\theta}_{v0}}\frac{\partial \theta'}{\partial z}.}_{bouyancy~term}
\end{equation}
This equation is elliptic - meaning that $p'$ at position $x_i$ requires knowledge of the flow field and temperature across the entire flow domain. Hence, local closure models that represent $\Pi_{ij}$ as a function of only local velocity statistics (or their gradients) at $x_i$ cannot accommodate the non-local effects of velocity and temperature at distant points. Nonetheless, Eq. \ref{eq:p_Poisson} underscores three mechanisms that historically formed the basis for modelling $\Pi_{ij}$. The first is known as the `rapid term' because it involves direct interaction between the mean strain rate ($\partial \overline{U_i}/\partial x_j$) and turbulence. The second is known as the `slow term' because it involves turbulent stresses that adjust after the mean strain rate has reacted to changes in boundary conditions. The third term is known as the buoyancy pressure term \citep{GibsonLaunder1978,katul1996inactive}. This term does not respond instantly to changes in the mean shear; however, it does respond instantly to changes in temperature gradients caused by density fluctuations, and thus shares some similarities with the rapid term. Spatial integration of the Poisson equation yields a fourth term - the spatial variations of the boundary conditions. These boundary conditions encode the so-called wall-blocking effect on the pressure-strain interaction \citep{launder1975progress}.

In operational closure schemes \citep{hanjalic1972reynolds,launder1975progress,mellor1982development,Heinze2016,hanjalic2021reassessment} the pressure-strain term $\Pi_{ij}^s$, by and large, follows this decomposition of Eq. \ref{eq:p_Poisson}
\begin{equation}
\label{eq:pressure_strain}
   \Pi_{ij}^s = \Pi_{ij}^T +\Pi_{ij}^S + \Pi_{ij}^B
\end{equation}
into the slow part represented by a linear return-to-isotropy $\Pi_{ij}^T$ and a rapid part that consists of the isotropization of production $\Pi_{ij}^S$ (shear and vorticity terms), as well as buoyancy $\Pi_{ij}^B$ (usually absorbed in the isotropization of the production).
The Rotta model \citep{Rotta1951} is then routinely used to close the slow $\Pi_{ij}^T$, following some revisions. 

The pressure-strain parametrizations $\Pi_{ij}^s$ (e.g., the Rotta closure scheme and corollary modifications) are mostly based on the decomposition of the strain rate tensor ${\partial\overline{U}_i}/{\partial x_j}$ into a symmetric part ($S_{ij}$) corresponding to shear (i.e., strain rate), and an anti-symmetric part ($R_{ij}$) corresponding to vorticity, given as
\begin{equation}
\frac{\partial\overline{U}_i}{\partial x_j} = S_{ij} + R_{ij},
\end{equation}
defined as 
\begin{equation}
\label{eq:IPshear}
S_{ij} = \frac{1}{2} \left[ \frac{\partial\overline{U}_i}{\partial x_j} + \frac{\partial\overline{U}_j}{\partial x_i}\right],
\end{equation}
and
\begin{equation}
\label{eq:IPvorticity}
R_{ij} = \frac{1}{2} \left[ \frac{\partial\overline{U}_i}{\partial x_j} - \frac{\partial\overline{U}_j}{\partial x_i}\right].
\end{equation}
Some mismatch of how the different parametrizations are applied exists. Some studies \citep{Zeman1981, Canuto2001} apply the parametrizations (Eq. \ref{eq:pressure_strain}) to the total $\Pi_{ij}$, while others \citep{Heinze2016} selectively apply it to the pressure-strain interaction term ($\Pi_{ij}^s$) only. When the pressure transport term (i.e. $\Pi_{ij}^t$) is negligible \citep{Stull88}, then the last term in Eq. \ref{eq:Reynolds} reduced to $\Pi_{ij}=\Pi_{ij}^s$.

Here, the approach of \citet{Heinze2016} is employed for pressure-strain parametrization. In this approach, the slow part of $\Pi_{ij}^s$ is due to the turbulence-turbulence interactions and is parameterized using the linear Rotta closure
\begin{equation}
\label{eq:Rotta}
\Pi_{ij}^T= -2\frac{c}{\tau} b_{ij}e.
\end{equation}
Here, $\tau = e/\mathrm{\varepsilon}$ is the relaxation time scale, $e=\frac{1}{2}\left( \overline{u'_k u'_k}\right)$ is the TKE, and $\mathrm{\varepsilon}$ is the mean TKE dissipation rate. Conceptually, $\tau$ measures the time it takes for $u'_k$ to de-correlate from itself (i.e., $u'_k$), which is why $\tau$ must reflect the slower integral scale (bottleneck in this decorrelation) instead of the faster micro-scales. Furthermore, $b_{ij}$ represents the anisotropy tensor and is the deviatoric part of the normalized Reynolds stress tensor
\begin{equation}
\label{eq:bij}
b_{ij} = \frac{\overline{u_i'u_j'}}{e} - \frac{2}{3} \delta_{ij}.
\end{equation}
Invariants of the anisotropy tensor have long been used to quantify the degree and type of anisotropy of the flow \citep{LumleyNewman1977,Pope2000}. In particular, the third invariant of the anisotropy tensor, related to the smallest eigenvalue $\lambda_{3}$ of the anisotropy tensor (Eq. \ref{eq:bij}) and defined as 
\begin{equation}
\label{eq:yB}
 y_{B}=(\sqrt{3}/2)(3 \lambda_{3} + 1),
\end{equation}
carries the information on the degree of anisotropy. Here the barycentric representation \citep{Banerjee2007} of the anisotropy invariant map \citep{LumleyNewman1977} is used.
 
The Rotta constant $c$ in Eq. \ref{eq:Rotta} has been predicted to take on values in the range of $c = 1.5 - 1.8$ \citep{Heinze2016} or $1 - 3$ \citep{Zeman1981}; for stable stratification, \citet[][note that they use half-variance and therefore the values have been updated here for full variance]{BouZeid2018} showed that it must be constrained between 1 and 5, but they use a value of 1.8 to compare to LES and DNS under stable and unstable conditions. 

The rapid terms in $\Pi_{ij}^s$ that form the isotropization of production due to shear (Eq. \ref{eq:IPshear}) and vorticity (Eq. \ref{eq:IPvorticity}) have been parametrized as a function of $e$ only
\begin{align} 
\label{eq:rapid_S}
\Pi_{ij}^S = & \frac{4}{5}S_{ij}e + C_{S1}^u(b_{jk}S_{ik} + b_{ik}S_{jk} -\frac{2}{3}b_{kl}S_{kl}\delta_{ij})e \\
& + C_{S2}^u(b_{ik}R_{jk} + b_{jk}R_{ik})e \nonumber,
\end{align}
where typical values of the two constants are $(C_{S1}^u, C_{S2}^u) = (\frac{12}{7},0)$ and $(\frac{3}{5},\frac{3}{5})$ (see table \ref{tab:tableClosureSchemes} in Appendix \ref{appendixClosureSchemes}). \cite{launder1975progress} and others \citep{so1977model} point to the particularly strong influence of the vorticity term (related to $R_{ij}$ in Eq. \ref{eq:rapid_S}) over curved surfaces. In the budgets of normal stresses in streamline coordinates, the $(4/5) S_{ij} e$ term is identically zero. 

In parallel, the rapid buoyancy term is parametrized as 
\begin{align}
\label{eq:rapid_B}
\Pi_{ij}^B= -C_{B}^u\frac{g}{\overline{\theta}}(\overline{u_j'\theta'}\delta_{i3} + \overline{u_{i}'\theta'}\delta_{j3}- \frac{2}{3}\overline{u_{k}'\theta'}\delta_{k3}),
\end{align}
where $C_{B}^u=\frac{3}{10} - \frac{3}{5}$, and the $\frac{3}{10}$ value corresponds to isotropic turbulence. 

Last, the wall blocking effects on the pressure-strain correlation would also need to be considered. One way to model this term assumes it to be proportional to $l_{wall}/z$, where $l_{wall}$ is presumed to scale with $e^{3/2}/\varepsilon$ \citep{GibsonLaunder1978}. As $z$ becomes large (larger than the integral scale), the effects of wall-blocking become small. These scaling arguments have, however, been developed without considering the effects of thermal stratification on wall-blocking.

Although the remaining terms in the budgets (Eq. \ref{eq:Reynolds}) also require closure assumptions in numerical models, in the observational datasets it is possible to assess some of them directly. This is particularly the case for the vertical turbulence transport terms ($T$), which can be estimated from multilevel towers. 

Finally, the Coriolis term ($Co$) is commonly neglected in surface-layer studies due to its low magnitude near the ground \citep[e.g.,][]{Stull88,KaimalFinnigan1994}.

\subsection{Reduced model for energy anisotropy}
\label{sec:Model_Reduced}


\begin{table}
\caption{The different versions of the models tested}
\label{tab:tableModels}
\centering
\scriptsize
\begin{tabular}{p{1.2cm}|p{1.7cm}|p{1.7cm}|p{1.7cm}|p{1.5cm}|p{4cm} }\toprule
 \textbf{Model Name}  & \textbf{Equations} & \textbf{Balanced Production and Dissipation} & \textbf{Rotta Closure}  & \textbf{Transport Terms} & \textbf{Rapid Terms} \\\midrule
 
  \textbf{R}  & \ref{eq:Reduced} & yes &  linear & no & no\\
 \textbf{Ra} & \ref{eq:Reduced_anis} & yes & adjusted & no & no\\
 \textbf{E} & \ref{eq:FullX}  - \ref{eq:FullZ} & no &  adjusted & no & yes \\
 \textbf{Et} & \ref{eq:FullX}  - \ref{eq:FullZ} & no &  adjusted & $T$ & yes \\
 \textbf{Etp} & \ref{eq:FullX}  - \ref{eq:FullZ} & no &  adjusted & $T$, $\Pi_{ij}^t$ & yes \\
 \midrule
 \textbf{Variants} & &  &   &  &  \\
 \midrule
\textbf{$_n$} & \ref{eq:Rotta_nonlin} &  &  non-linear &  &   \\
\textbf{$_1$} &  &  &   &  &  $C_{B}^u = C_{S1}^u = C_{S2}^u = 0$\\
\textbf{$_2$} &  &  &   &  &  $C_{B}^u = 0.3, C_{S1}^u = C_{S2}^u = 0.6$\\
\textbf{$_3$} &  &  &   &  &  $C_{B}^u = C_{S1}^u = C_{S2}^u = 0.6$\\
\textbf{$_4$} &  &  &   &  &  $C_{B}^u = 0.6, C_{S1}^u = 12/7, C_{S2}^u = 0$\\

\bottomrule
 
\end{tabular}
\end{table}

While an important contribution to the anisotropy of the flow stems from the off-diagonal Reynolds stress tensor terms, i.e., momentum fluxes \citep[cf.][]{stiperski2021convective}, the focus here is on the anisotropy of the normal Reynolds stresses (i.e., velocity variances), as they provide information on the anisotropy in energy distribution. As a starting point, a reduced model for the Reynolds stresses \citep{BouZeid2018} is employed to examine the evolution of anisotropy with increasing instability. This model for half-variances assumes stationary, planar homogeneous conditions with no subsidence (all horizontal terms, as well as mean vertical advection are zero), with isotropic dissipation, negligible contribution from both turbulent ($T$) and pressure ($\Pi_{ij}^t$) transport terms, and pressure-strain correlations parametrized using only the standard Rotta closure (Eq. \ref{eq:Rotta}) without the rapid terms related to isotropization of the production. The latter allows writing the half-variance budgets in terms of velocity variance ratios ($\overline{u_i'^2}/e$), which are functions of the TKE generating mechanisms - mechanical production ($S_m = -\overline{u'w'}\partial{\overline{U}}/\partial z$) and buoyancy production or damping ($B = \overline{w'\theta_v'}g/\overline{\theta}_{v0}$) only. 
These component-wise velocity variance budgets are given by
\begin{align}
\frac{\overline{u'^2}}{e} & = \frac{S_m}{c \varepsilon}\left( \frac{2}{3} + \frac{1}{3} R_{if} \right) + \frac{2}{3}, \nonumber \\
\frac{\overline{v'^2}}{e} & = \frac{S_m}{c \varepsilon}\left( -\frac{1}{3} + \frac{1}{3} R_{if} \right) + \frac{2}{3}, \label{eq:Reduced_full} \\
\frac{\overline{w'^2}}{e} & = \frac{S_m}{c \varepsilon}\left( -\frac{1}{3} - \frac{2}{3} R_{if} \right) + \frac{2}{3}, \nonumber
\end{align}
where $R_{if}=-B/S_m$ is the flux Richardson number, quantifying the relative importance of buoyancy forces over the mechanical production of TKE. 
Note that here, the Rotta constant $c$ should attain values that are half as small as the ones reported in the previous section. 
The sum of the three components yields the TKE budget, which under the same assumptions can be simplified to a balance between molecular dissipation ($\varepsilon$), mechanical production ($S_m$), and buoyancy (production or damping) ($B$) following
\begin{equation}
\label{eq:shear}
\frac{S_m}{\varepsilon} = \frac{1}{1 - R_{if}},
\end{equation}
which allows the variance ratios to be expressed as a function of $R_{if}$ only \citep{BouZeid2018}. The full set of simplified equations are then given by

\begin{align}
\frac{\overline{u'^2}}{e} & = \frac{1}{3c }\left( \frac{2 + R_{if}}{1-R_{if}} \right) + \frac{2}{3}, \nonumber \\
\frac{\overline{v'^2}}{e} & = \frac{1}{3c }\left( \frac{-1 + R_{if}}{1-R_{if}}\right) + \frac{2}{3}=\frac{2}{3}\left(1-\frac{1}{2 c}\right), \label{eq:Reduced} \\
\frac{\overline{w'^2}}{e} & = \frac{1}{3c}\left( \frac{-1-2R_{if}}{1-R_{if}} \right) + \frac{2}{3}, \nonumber
\end{align}
We refer to this model as \textbf{Model R} (see table \ref{tab:tableModels}) and point out that for this model, $\overline{v'^2}/{e}$ is constant and independent of $R_{if}$. In both versions of this reduced model (Eq. \ref{eq:Reduced_full} and Eq. \ref{eq:Reduced}), contribution of each variance to the total $e$ is a result of changing importance of production terms quantified through stratification ($R_{if}$) on the one hand, and the pressure redistribution through which the less-energetic components receive energy from the energetic components at an equal rate, irrespective of stratification.  This reduced model serves as a reference for assessing contributions from other anisotropy-generating mechanisms.

\begin{figure}
    \centering
        \includegraphics[width = 0.7\linewidth]{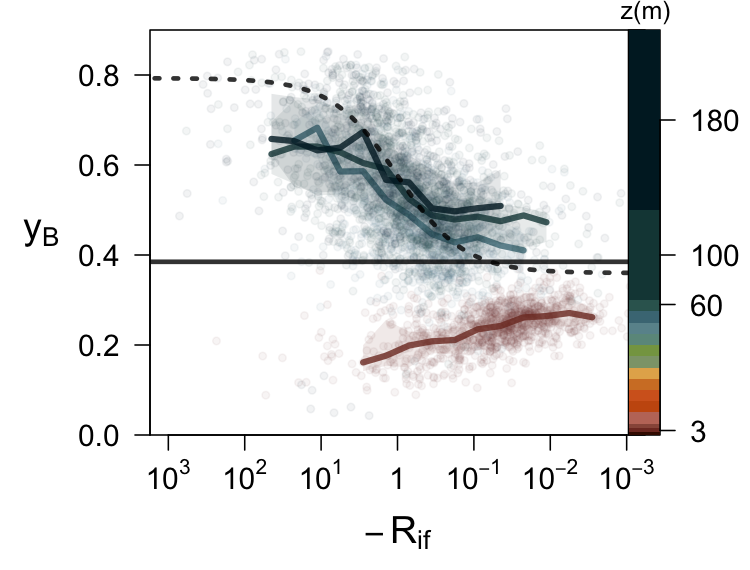}
    \caption{Degree of energy anisotropy $y_{B}$ as a function of flux Richardson number $R_{if}$ for the Cabauw tower data. Coloured points represent individual averaging periods for the four measurement heights, where the full coloured lines are the bin averages computed at logarithmically spaced $R_{if}$ and shading is the interquartile range. Solid black line corresponds to the predictions of reduced \textbf{Model R} (Eq. \ref{eq:Reduced} with $c = 0.9$), while the dashed black curve corresponds to the prediction of the reduced model with adjusted Rotta constant ($c = 3$) and added wall-blocking $[a_u,a_v,a_w] = [1.25, 1.15, 0.58]$ (\textbf{Model Ra}).}
    \label{fig:Cabauw_anis}
\end{figure}

Without a contribution of momentum fluxes, the smallest component of $b_{ij}$, corresponding to the smallest variance ratio, equals the smallest eigenvalue. Without a source of its own (shear or buoyancy) in the reduced budget, the smallest variance ratio in the $R_{if}<0$ range is always the spanwise variance $\overline{v'^2}/e$ \citep[see Fig. 3 in][]{BouZeid2018}, whose equation is independent of $R_{if}$ (Eq. \ref{eq:Reduced}). As a result, the degree of anisotropy based on the smallest eigenvalue (as defined in Eq. \ref{eq:yB}) is constant and is not a function of $R_{if}$ (Fig. \ref{fig:Cabauw_anis}) for the reduced model. The data in Fig. \ref{fig:Cabauw_anis}, however, suggest a significant influence of both stratification and height on the degree anisotropy $y_{B}$.

\subsection{Revisions to the reduced model for energy anisotropy}
\label{sec:Model_Extended}

The model in Eq. \ref{eq:Reduced} can be expanded in multiple ways to accommodate different sources of anisotropy. A process missing from the classic Rotta model is the existence of wall effects (or `pressure echo') due to the rigid boundary \citep{Pope2000,mccoll2016mean}, disproportionately affecting the wall-normal variance $\overline{w'^2}$. This source of anisotropy can be added to the Rotta closure itself by allowing the energetics of individual velocity components not to strive towards a state of equipartition. Instead, a state that preserves a certain degree of anisotropy is proposed, where the modified Rotta closure can be expressed as
\begin{equation}
\Pi_{ij}^T = -\frac{c\varepsilon}{e}\left(\overline{u'_{i}u'_{j}} - \frac{2}{3}e \delta_{ij} a_i \right).
\label{eq:Rotta_anis}
\end{equation}
The coefficients $a_i$ (or $a_u$, $a_v$, and $a_w$) satisfy $a_1 + a_2 + a_3 = 3$ to allow the summed redistribution terms not to produce or dissipate TKE. This is the essence of the wall-function corrections to the pressure-strain interaction. The variance equations then become
\begin{align}
\frac{\overline{u'^2}}{e} & = \frac{1}{3c }\left( \frac{2 + R_{if}}{1-R_{if}} \right) + \frac{2}{3}a_u, \nonumber \\
\frac{\overline{v'^2}}{e} & = \frac{1}{3c }\left( \frac{-1 + R_{if}}{1-R_{if}}\right) + \frac{2}{3}a_v= -\frac{1}{3c } + \frac{2}{3}a_v, \label{eq:Reduced_anis} \\
\frac{\overline{w'^2}}{e} & = \frac{1}{3c}\left( \frac{-1-2R_{if}}{1-R_{if}} \right) + \frac{2}{3}a_w, \nonumber
\end{align}
We refer to this model as \textbf{Model Ra} (see table \ref{tab:tableModels}). Note that,  if $a_w$ is sufficiently small and thus $\overline{w'_2}/e$ becomes the smallest variance ratio, this model allows the degree of anisotropy to vary with stratification (see Fig. \ref{fig:Cabauw_anis}, dashed line). 

Beyond wall-blocking effects, another source of anisotropy in the ASL is the anisotropy of TKE dissipation rate, generally assumed to occur at micro-scales where the eddies are isotropic in high Reynolds number flows \citep{Kolmogorov1941a}. This assumption is employed in the reduced model. However, this assumption is not guaranteed in realistic ASL flows \citep[e.g.,][]{Biltoft2001}.
The inclusion of dissipation anisotropy would modify the constants in the first term on the right-hand side of Eq. \ref{eq:Reduced_anis} (cf. Appendix \ref{appendixAnisotropicDissipation}), and would thus only affect the magnitude of the Rotta constant. To accommodate this effect, the Rotta constants $c_i$ are here allowed to vary between the velocity components as suggested by previous studies \citep[see][]{Heinze2016,Yi2025}. 

Next, turbulent transport terms $T_i$, known to be relevant in buoyancy-driven conditions \citep[e.g.,][]{stiperski2021convective}, can be added to the reduced model \citep[][cf. their Appendix A]{BouZeid2018}. Allowing for transport in this modelling framework was shown to be critical for explaining turbulence levels under stable conditions at heights where the local dissipation plus buoyant destruction exceeded shear production \citep{Freire2019}. Under the assumption of planar homogeneity, only the vertical turbulence transport terms are relevant for extending the reduced model. In case transport is included, however, the balance between production and dissipation mechanisms encoded in Eq. \ref{eq:shear} no longer holds, and the full form of the budget equations (Eq. \ref{eq:Reduced_full}) must be used, where the dissipation is maintained to be isotropic.

Finally, the role of the rapid return-to-isotropy terms neglected in the reduced model can also be added. These include the isotropization of production terms due to shear, vorticity and buoyancy (Eqs. \ref{eq:rapid_S}-\ref{eq:rapid_B}, table \ref{tab:tableClosureSchemes} in Appendix \ref{appendixClosureSchemes}). The final model accommodating all these revisions is given by

\begin{align}
  \frac{\overline{u'^2}}{e}= \nonumber \\& \frac{1}{c_u \varepsilon} \left[ \left( \frac{2}{3} + \frac{1}{3} R_{if} \right)S_m + \frac{1}{3}C_{B}^uB - \left(\frac{1}{3} C^u_{S1} + C^u_{S2} \right)\frac{S_m}{2} +  \frac{\left( T_w + T_v -2T_u \right)}{3} \right]+ \frac{2}{3}a_u, \label{eq:FullX} \\
  \frac{\overline{v'^2}}{e}= \nonumber \\ & \frac{1}{c_v \varepsilon} \left[ \left( -\frac{1}{3} + \frac{1}{3} R_{if} \right)S_m + \frac{1}{3}C_{B}^uB + \frac{2}{3}C^u_{S1}\frac{S_m}{2} + \frac{\left(T_w +T_u -2T_v\right)}{3} \right] + \frac{2}{3}a_v, 
  \label{eq:FullY} \\
  \frac{\overline{w'^2}}{e}= \nonumber \\ & \frac{1}{c_w \varepsilon} \left[ \left( -\frac{1}{3} - \frac{2}{3} R_{if} \right)S_m - \frac{2}{3}C_{B}^u B - \left(\frac{1}{3} C^u_{S1} - C^u_{S2} \right)\frac{S_m}{2} + \frac{\left( T_u + T_v -2T_w \right)}{3} \right] + \frac{2}{3}a_w. \label{eq:FullZ} 
\end{align}

Here $T_u$ (or $T_1$), $T_v$ (or $T_2$) and $T_w$ (or $T_3$) are the vertical transport terms of the respective half variance ($T_i = -(1/2)\partial{\overline{w'u_i'^2}}/\partial{z}$). This model will be referred to as \textbf{Model E} with its several variations (see table \ref{tab:tableModels}). 

The final model has a number of parameters that need to be externally supplied. Unless otherwise specified, the rapid term constants are set to $C^u_{B} = C^u_{S1} = C^u_{S2} = 3/5$ \citep{Heinze2016}. The constant $c$ in the Rotta model has to be different in case the full \textbf{Model E} (Eqs \ref{eq:FullX} - \ref{eq:FullZ}) is used or if only the slow (i.e., Rotta) terms are kept (Eqs \ref{eq:Reduced} and \ref{eq:Reduced_anis}, \textbf{Models R} and \textbf{Ra}). In the full model, it was reported elsewhere \citep{Heinze2016} that values between 0.5 - 1.5 (and possibly centred at 1) are plausible (note that these values have been adjusted for half-variance budgets compared to those presented in Sect. \ref{sec:reynoldsstressbudget}); however, when approximating the return to isotropy using the slow-part only, $c=0.9$ \citep{BouZeid2018,Pope2000}.   

An earlier study noted that the contribution of the transport terms to the non-dimensional variances depends on whether the individual transport components act together or against each other \citep{BouZeid2018}. Expressing the transport of TKE as $T_e= T_u + T_v +T_w$, the transport contribution for a given component $i$ can be expressed as $T_e-3T_i$ (cf. Eqs. \ref{eq:FullX}-\ref{eq:FullZ}). Since the vertical turbulent transport in the near-surface region of the convective boundary layer commonly carries a negative sign ($e$ is exported from ASL into the mixed layer), these transport terms are all expected to be negative \citep[this indeed was found to hold except under very stable stratification by ][]{Freire2019}. If the component $i$ is the least energetic, it will most likely result in a weaker negative transport than the other terms, and thus $T_e-3T_i<0$ \citep[again supported by the results of ][that indicated that $T_w\approx 0.28Te$ ]{Freire2019}, while for the most energetic component, we expect $T_e-3T_i>0$. Therefore, the contribution of the net transport terms boosts the most energetic component relative to the least energetic, and acts against the return-to-isotropy process. 

Additionally, the vertical pressure transport term ($\Pi_{ww}^t$) can also be included into $T_w$ if its magnitude can be estimated. Most measurement campaigns, however, do not include barometers measuring at sufficient temporal resolution with an adequate frequency response, and therefore do not allow its direct estimation. The effect of this term on anisotropy is discussed later on in Sect. \ref{sec:Model_Extended}. 

\subsection{Non-linear Rotta closures}
\label{sec:non-linear}

Apart from the modifications introduced to the Rotta closure in the previous sub-section (cf. Eq. \ref{eq:Rotta_anis}), a number of alternative non-linear and anisotropic Rotta models exist that include the anisotropy invariants into the closure directly. Here, we explore the performance of the following formulation from \cite{Heinze2016}
\begin{equation}
\label{eq:Rotta_nonlin}
\Pi_{ij}^T = -\frac{C_{T1}^u\varepsilon}{e}\left[b_{ij} + C_{T2}^u\left(b_{ik}b_{jk} -\frac{1}{3}A_2\delta_{ij} \right) \right]e,
\end{equation}
where $C_{T1}^u = \left(3.75 A_2^{1/2} + 1 \right)A$, $C_{T2}^u = 0.7$. Here $A_2 = b_{ij}b_{ji}$ and $A_3 = b_{ij}b_{jk}b_{ki}$ are the invariants of $b_{ij}$ and form the the flatness parameter $A = 1 - {(9/8)}\left(A_2 - A_3 \right)$. 
This model will be referred to with the subscript $n$ (table \ref{tab:tableModels}).

\section{Data and methods}
\label{sec:data}


\begin{table}
\caption{Information on the datasets used in the study}
\label{tab:tableData}
\centering
\scriptsize
\begin{tabular}{p{2cm}|p{1.7cm}|p{1.8cm}|p{1.5cm}|p{1.8cm}|p{1.5cm}|p{1.5cm} }\toprule
 \textbf{Station Name}  & \textbf{Terrain} & \textbf{Measurement heights [m]}  &   \textbf{Resolution [Hz]} &\textbf{Surface Type} & \textbf{Roughness Length [m]} & \textbf{Data length}\\\midrule
 
  \textbf{AHATS}  & flat, horizontally homogeneous & 1.55, 3.3, 4.24, 5.53, 7.08, 8.05 &  20&fallow land & 0.021 & Jun--Aug 2008\\
 \textbf{Cabauw} & flat, weakly inhomogeneous & 3, 60, 100, 180 &  10&mixed agricultural & 0.025 &Jul--Oct 2007\\
  \textbf{METCRAX}  II & 1$^{\circ}$ slope, horizontally homogeneous & 5, 10, 15, 20, 25, 30, 35, 40, 45, 50 &  20&desert with shrubs & 0.053 & Oct 2013\\
\textbf{M2HATS}  & flat, horizontally homogeneous & 3, 4, 7, 15, 28 &  60 & desert with shrubs & 0.027 & 1 Aug - 16 Sept 2023\\ 
\bottomrule
 
\end{tabular}
\end{table}

\subsection{Datasets and turbulence data post-processing}
A number of datasets representative of mostly flat and horizontally homogeneous (i.e., canonical) terrain are employed. These datasets are the vertical tower in the AHATS \citep{Nguyen2013}
experiment, the Cabauw tower \citep{BeljaarsBosveld1997}, NEAR tower from the METCRAX II experiment \citep{Lehner2016}, and the combined profile from the two multilevel towers (t0) at M2HATS campaign. M2HATS stands out as the dataset where in addition to sonic anemometers, the nano-barometers (Digiquartz Paroscientific 6000 measuring at 20Hz) were installed at each observational level, allowing a direct estimation of the pressure transport term. Characteristics of individual datasets are summarized in table \ref{tab:tableData}. 

Turbulence time series from the different datasets and sites were processed using a uniform procedure described in prior studies \citep{Stiperski2018, Stiperski2019, stiperski2021convective, Stiperski2021, Stiperski2023}. Turbulence statistics were computed over 30\,min block averages, with prior linear de-trending. The 30\,min average is the standard processing time in atmospheric turbulence studies \citep[e.g.,][]{Aubinet2012}, and ensures that the largest convective eddies are represented in the temporal mean, while eliminating the non-turbulent signals and the influence of the daily cycle (i.e., non-stationarity). Using a 1\,h averaging time had no substantial influence on the results (not shown). 

Data were rotated into the streamline coordinates using a double rotation procedure \citep{Aubinet2012} in which the mean spanwise ($\overline{V}$) and wall-normal ($\overline{W}$) velocity component over each 30 min period are set to zero, and the streamwise direction ($\overline{U}$) is aligned with the mean wind direction.

Data were quality-controlled for instrument errors and for values outside of the physical range. Additionally, only periods for which the spectral slope of the streamwise and spanwise spectra in the inertial subrange equalled $-5/3$ with a $20\%$ error margin, and for which the stationarity of the mean wind speed was limited to $30\%$ based on the common stationarity test of \cite{FokenWichura1996}, were retained. No additional quality criteria were required (e.g., stationarity of higher-order statistics).

Finally, only the unstable daytime surface layer was explored. The buoyancy flux ($\overline{w'\theta_v'}$, where $\theta_v$ was taken to equal the sonic temperature) was therefore required to be positive at all observational heights. This criterion avoids classifying (evening/morning) transition periods, or nighttime countergradient fluxes as daytime unstably stratified turbulence. 

\subsection{Computation of the Reynolds budgets terms}

The Reynolds stress budget terms were evaluated at each observational height. The vertical wind shear as part of the shear production term ($S_m$) was obtained by fitting the function $\overline{U}(z) = a + b\,z + c\,\ln{z}$ through the mean velocity profile, and evaluating the gradient analytically. 
Vertical components of the turbulence transport terms $T_i = -(1/2)\partial{\overline{w'u_i'^2}}/\partial{z}$ were computed by fitting a polynomial through the triple correlation terms ($\overline{w'u_i'^2}$) and analytically evaluating the gradient. A second-order polynomial in $z$, a third-order polynomial in $z$, or a more complex polynomial functions of the form
$a + b\,z + c\,z^2 + d\,\ln{z}$ were explored, and the form that produced the smallest root mean square fitting error was chosen for each averaging period for gradient evaluations. Given the uncertainty in the behaviour of the transport terms, this fitting procedure is a source of non-negligible uncertainty in the estimated Reynolds stress budgets. 

The TKE dissipation rate ($\varepsilon$) was determined from the inertial subrange of the 30~min spectra of the streamwise velocity component, following the inertial dissipation method \citep{Chamecki2004}
\begin{equation}
S_u = \alpha_uk^{-5/3}\varepsilon^{2/3},
\end{equation}
where $S_u$ is the streamwise power spectral density, $k=(2\pi f)/\overline{U}$ is the streamwise wavenumber determined based on Taylor's hypothesis, $f$ the frequency, and $\alpha_u = 18/55$ Kolmogorov constants for streamwise component. To ensure that the TKE dissipation rate is representative of the inertial subrange and to avoid aliasing at high frequencies, the frequency range over which $\varepsilon$ was estimated was limited between the frequency corresponding to the peak in the pre-multiplied wall-normal velocity spectrum at frequencies corresponding to $k z>1$ on the one hand, and $f=10^{-0.1}$ on the other. Finally, the TKE dissipation rate for each averaging period was computed as the median of the dissipation rate estimates at each frequency within the inertial subrange. The spectra were smoothed prior to computing the dissipation rate by computing bin averages of spectral density within logarithmically spaced bins in the frequency space. Additionally, the spectral slope of the inertial subrange was estimated from the smoothed spectra using robust linear regression in log-log space, and used as a quality criterion for the computed TKE dissipation. 

It has been known for quite some time now that while the scaling laws (i.e., $k^{-5/3}$) are robust to the local isotropy assumption, the spectral ratios that determine the Kolmogorov constants for each velocity component, are not \citep{saddoughi1994local,hsieh1997dissipation} as they may be impacted by turbulent intensity and intermittency, thus requiring corrections. The estimated dissipation rates were therefore multiplied by the turbulence intensity correction ($F_u = 1+(11/9)I_u^2$), where $I_u = \sqrt{\overline{u'^2}}/\overline{U}$ is the turbulence intensity.

The integral length scales of the spanwise and wall-normal velocity components ($\lambda_u,\lambda_w$) were estimates from the auto-correlation function of the respective velocity components. Here, first the integral time scale was determined as the time scale at which the autocorrelation function drops to $1/exp$, where $exp$ is the base of the natural logarithm, and then subsequently converted to integral length scale through Taylor's hypothesis.

\begin{figure}
    \centering
        \includegraphics[width = 0.6\linewidth]{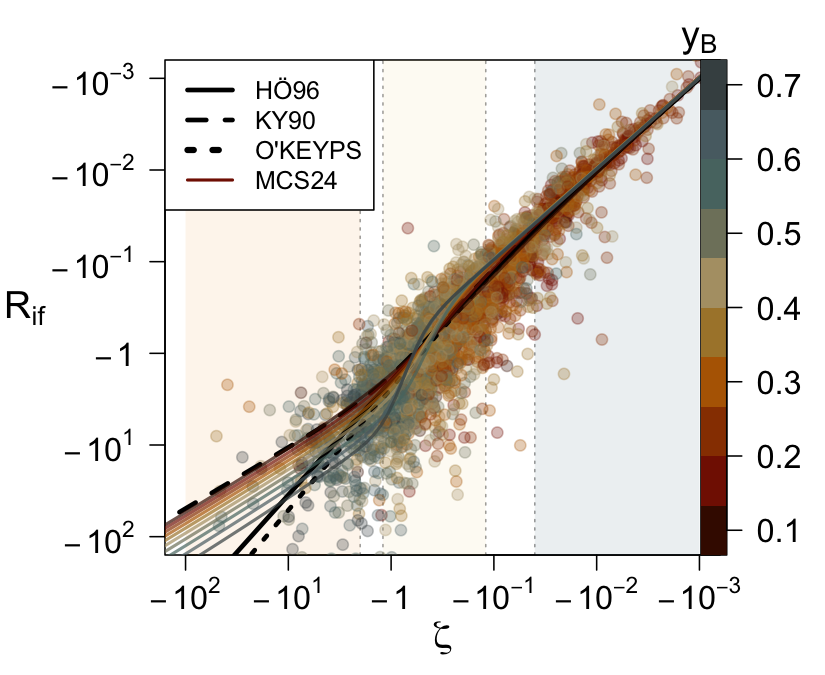}
    \caption{Relation between the local stability parameter $\zeta=z/\Lambda$ and the flux Richardson number $R_{if}$ for the Cabauw tower as a function of the degree of anisotropy $y_B$ (colour). Dots correspond to observational averaging periods. Curves correspond to the different scaling relations for $\Phi_m$: full black curve \cite{Hogstrom96}, dashed black curve \cite{KaderYaglom1990}, dotted black curve the O'KEYPS equation \cite{LumleyPanofsky1964}, and coloured full curve \cite{Mosso2024} that includes the degree of anisotropy into scaling (colour of the curves). Vertical coloured ranges separated by thin dotted lines correspond to the three subranges of \cite{KaderYaglom1990}: dynamic (blue, $-\zeta < 0.04$), dynamic-convective (yellow, $-\zeta =[0.12 - 1.2]$), and convective (orange, $-\zeta > 2$)}. 
    \label{fig:Cabauw_RiZL}
\end{figure}

In the ASL turbulence literature, the strength of thermal stratification is quantified based on either $R_{if}$ or the atmospheric stability parameter $\zeta$. 
The $R_{if}$ was computed from the turbulent fluxes and gradients at each height as
\begin{equation}
    R_{if} = \frac{g}{\overline{\theta}_{v0}} \frac{\overline{w
'\theta_v'}}{\overline{u'w'} \frac{\partial \overline{U}}{\partial z} }, 
\end{equation}
where $g$ is the gravitational acceleration, $\overline{u'w'}$ is the momentum flux, $\overline{w'\theta_v'}$ the wall-normal buoyancy flux, and $\theta_{v}$ the virtual potential temperature assumed to equal the sonic anemometer temperature, while $\overline{\theta}_{v0}$ is the virtual potential temperature of the background state. The local stability parameter is defined as $\zeta = z/\Lambda$, where $\Lambda$ is the Obukhov length 
\begin{equation}
    \Lambda = - \frac{u_*^3}{(\overline{w
'\theta_v'}/\overline{\theta}_{v0})\kappa g}, 
\end{equation}
$\kappa$ is the von K{\'a}rm{\'a}n constant set to $0.4$, and $u_* = (\overline{u'w'}^2 + \overline{v'w'}^2)^{1/4}$ is the local friction velocity accounting for both the frictional stress ($\overline{u'w'}$) and directional stress ($\overline{v'w'}$).  The relation between $\zeta$ and $R_{if}$ is shown in Fig. \ref{fig:Cabauw_RiZL} to facilitate delineation of the various sublayers within the ASL that are routinely defined based on $\zeta$ instead of $R_{if}$ used here.
As expected, the local $R_{if}$ is non-linearly related to $\zeta$ and this relation is given by  
\begin{equation}
    R_{if}=\frac{\zeta}{\phi_m(\zeta)}, 
\end{equation}
where $\phi_m(\zeta)$ is the stability function for the mean velocity gradient \citep{Stull88}, with $\phi_m(0)=1$ recovering the neutral law-of-the wall. Multiple stability functions proposed over the decades \citep{Hogstrom96,KaderYaglom1990,LumleyPanofsky1964,Mosso2024}, are illustrated in Fig. \ref{fig:Cabauw_RiZL}. Recently proposed scaling relations of \citet{Mosso2024} incorporate the degree of anisotropy $y_B$ into the scaling relations of \cite{KaderYaglom1990}. 
 
\subsection{Statistical analysis}

In the explored models (cf. table \ref{tab:tableModels}), the Rotta constants $c_i$ and wall factors $a_i$ were left as free parameters to allow the model to match data. They were estimated from the orthogonal distance regression (also known as total least squares) of the model against the observed variance ratios, as this method accounts for uncertainty in both the predictor and the response. Given the large non-linearity of the data as a function of $R_{if}$ at low observational heights, bin averages of the variance ratios and the model on logarithmically spaced $R_{if}$ were computed before the regression analysis was performed. Since the best model and the variance ratios should be linearly related, we tested how well the model captured this relation through the Pearson correlation coefficient, ignoring the fitted Rotta constant. Thus, the correlation coefficient, as well as the Rotta constant were allowed to attain negative values. When the Rotta constant obtained by the fitting procedure was negative, it was artificially set to $c = 1.8$ in Fig. \ref{fig:Cabauw_Model} for visualization purposes only. 

\section{Drivers of energy anisotropy}
\label{sec:drivers}

\subsection{TKE budget in the ASL}
\label{sec:TKE}
\begin{figure}
    \centering
        \includegraphics[width = 0.8\linewidth]{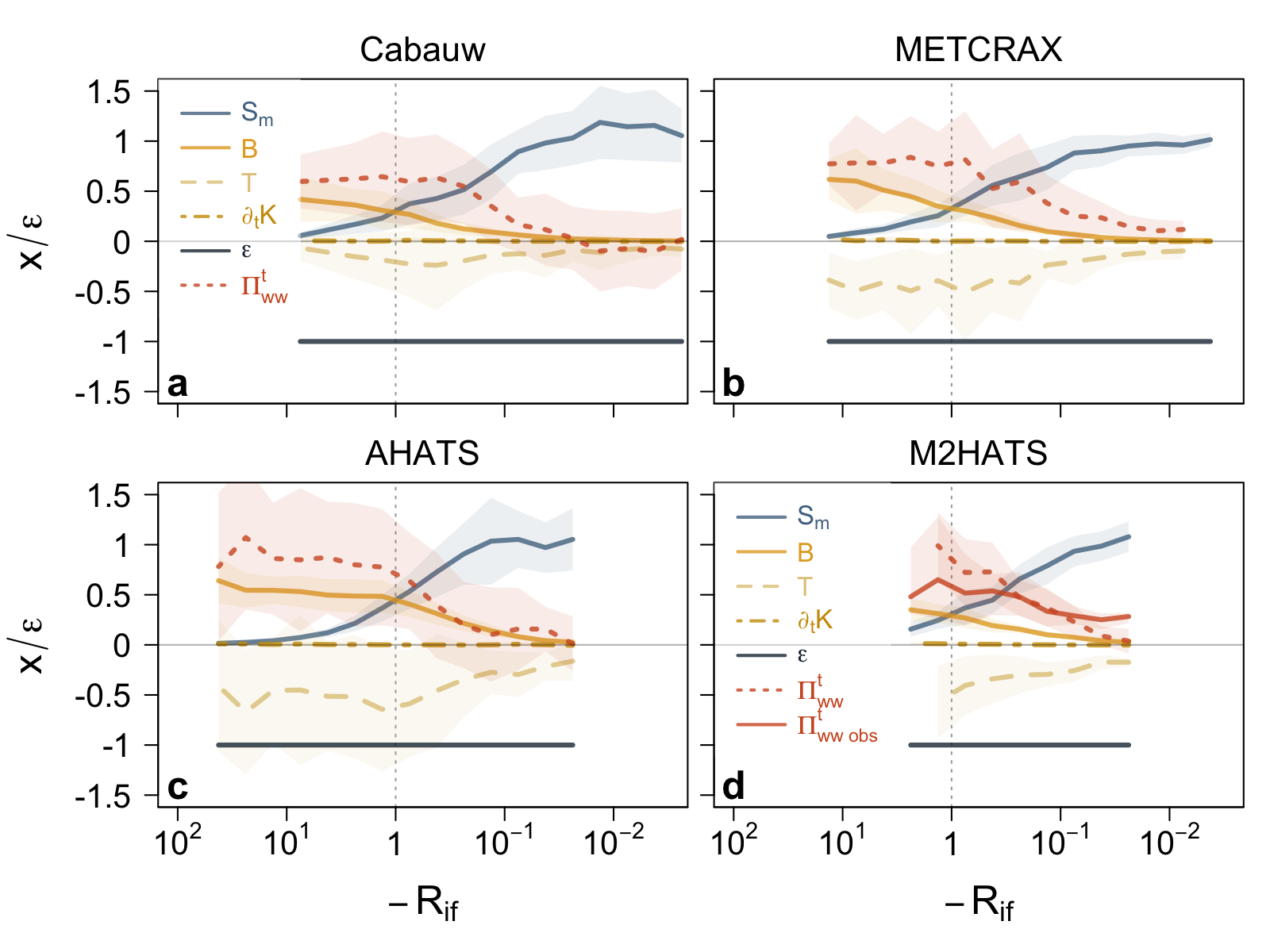}
    \caption{Terms of the TKE budget (colour) normalized by the dissipation rate as a function of $R_{if}$ for the average of 3 to 10~m levels at Cabauw, METCRAX, AHATS, and M2HATS towers. Lines correspond to logarithmically spaced bin averages, while the shading is the interquartile range. For variable names, see Eq. \ref{eq:Reynolds}. Note the significance of the vertical pressure transport $\Pi^t_{ww}$ with decreasing $R_{if}$ at all sites.}
    \label{fig:Flat_TKE}
\end{figure}

Before evaluating the Reynolds stress budgets and attendant simplifications, it is necessary to test the closure of the TKE budget itself. The TKE budget terms for the $z$=3 m measurement height at all the towers (Fig. \ref{fig:Flat_TKE}) show the expected behaviour, with the dominance of shear production ($S_m$) almost balanced by dissipation ($\varepsilon$) in the near-neutral stratification, and the rising importance of buoyancy ($B$) with increasing instability. On the other hand, the vertical turbulent transport term ($T$, yellow dashed line in Fig. \ref{fig:Flat_TKE}) is non-zero outside of neutral stratification, and has a magnitude that is commensurate or even exceeds the buoyancy production term ($B$) in majority of the datasets. Nonetheless, even with the addition of turbulent transport, the budget is not closed with the terms that we can directly compute. In fact, the observed residual (dashed red line in Fig. \ref{fig:Flat_TKE}, labelled as $\Pi^t_{ww}$) exceeds both the vertical turbulent transport term, as well as the buoyancy production. 

Given that the datasets were collected over nominally homogeneous conditions, horizontal terms \citep[advection, flux divergences, as well as horizontal shear production, cf.,][]{Goger2018} are small and unlikely contributors to the observed imbalance. This imbalance in the budget is therefore likely to stem from the vertical pressure transport term $\Pi^t_{ww}$ \citep[see][]{Wyngaard2010}. Although not routinely measured, LES studies have already highlighted that the pressure transport term is non-negligible in the convective boundary layer over flat and horizontally homogeneous terrain \citep[e.g.,][]{Wyngaardetal1971,Wyngaard1973,MoengSullivan1994,Lin2000,NguyenTong2015,Ding2018}, and that its sign is positive (the same as the observed imbalance). Following \cite{Wyngaard2010}, we therefore attribute the residual of the TKE budget to the vertical pressure transport ($\Pi_{ww}^t$) and treat it as such in the rest of the manuscript. 
The results (dashed red line in Fig. \ref{fig:Flat_TKE}) show that this estimated pressure transport term for 3 m level is near-zero in near-neutral stratification, and positive and on the order of magnitude of $B$ in the convective range, as previously observed \citep{Wyngaard2010,RotachHoltslag2025}. Thus, our estimate agrees with the results of LES despite the limited resolution of LES at heights probed by the observations. 

Direct measurement of turbulent pressure at each observational height in the M2HATS dataset provides a direct test of the assumption that the TKE budget imbalance stems from vertical pressure transport (Fig. \ref{fig:Flat_TKE}d). The observed pressure transport and the one estimated as the residual of the TKE budget (full and dashed red lines in Fig. \ref{fig:Flat_TKE}d) show acceptable agreement in terms of sign, order of magnitude, and increasing tendency with increasing $-R_{if}$. 
The TKE budget closure, therefore, suggests that the vertical pressure transport term needs to be accounted for when evaluating the \textbf{Model E} (version \textbf{Etp}). 

Despite the acceptable agreement between the estimated and measured pressure transport (only available at one of the sites), some uncertainties in the budget remain. One source is the uncertainty associated with estimating turbulence transport. The computation of triple correlations is in itself uncertain, as longer averages are usually required for the convergence of higher order terms \citep{Wyngaard1973,lenschow1994long,saddoughi1994local,huang2022profiles}, whereas the functional shape of triple correlations with height needed for analytically estimating gradients is also not well known \citep{Wyngaardetal1971}. The same uncertainty in analytically estimating the pressure transport term is also present. Finally, some uncertainty is associated with the computation of the TKE dissipation rates from the inertial subrange due to competing influences of path-averaging and signal aliasing of sonic anemometers \citep[see,][]{Chamecki2004,Freire2019}. These uncertainties restrict the subsequent evaluation of the Reynolds stress budgets to an assessment of the plausible role of different processes in the individual variance budget equations, but allow a focus on the processes captured by the aforementioned anisotropy models. 

\subsection{Performance of the reduced model for the Reynolds stresses}
\label{sec:simple}

\begin{figure}
    \centering
        \includegraphics[width = 1\linewidth]{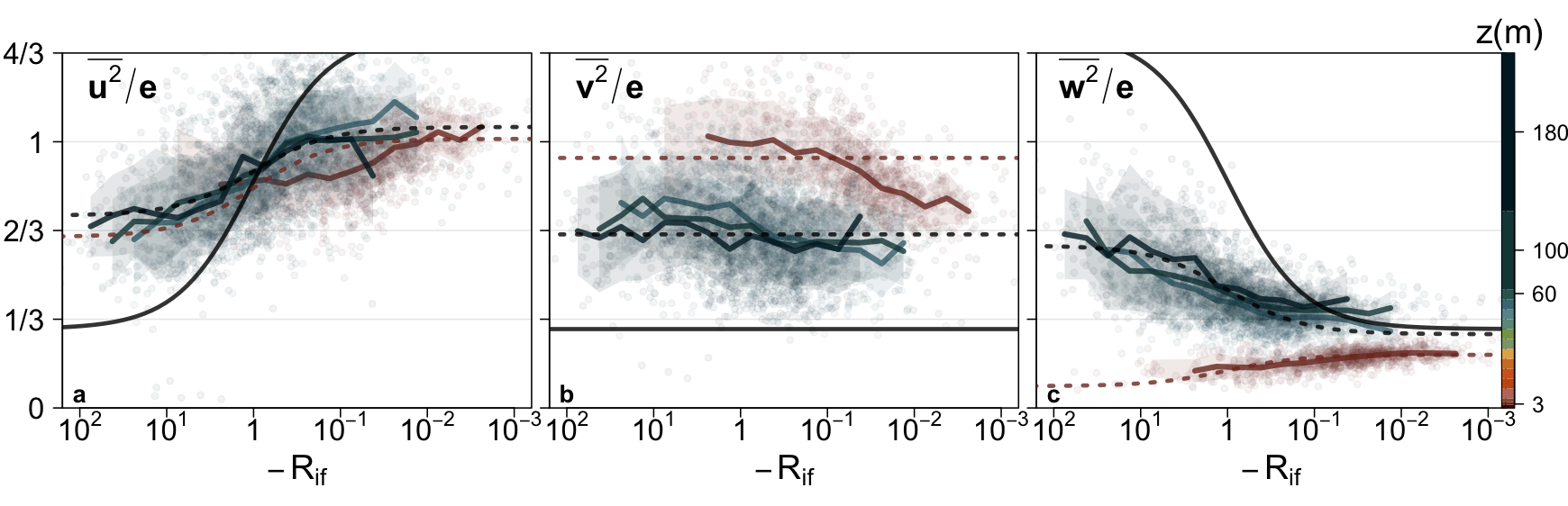}
    \caption{Bin averages of a) streamwise $\overline{u'^2}/e$, b) spanwise $\overline{v'^2}/e$, and c) wall-normal $\overline{w'^2}/e$ velocity variance ratio as a function of $R_{if}$ for the Cabauw tower. Four measurement heights are shown in colours. Full lines show bin averages computed at the logarithmically spaced $R_{if}$, while shading corresponds to the interquartile range. Full black curve corresponds to the predictions of reduced \textbf{Model R} (Eq. \ref{eq:Reduced}) with $c = 0.9$, while the dashed curves correspond to the predictions of the reduced \textbf{Model Ra} with an adjusted Rotta constant $c = 3$ and wall-blocking added (Eq. \ref{eq:Reduced_anis}). Here the wall-blocking and Rotta constants were obtained from a robust linear fit for the upper levels (60 - 180\,m, black) and first level (3\,m, brown) separately.}
    \label{fig:Cabauw_var}
\end{figure}

The individual variance ratios (i.e. as fractions of TKE) for the Cabauw tower are first explored because this tower's height (highest measurement level at 180m) allows probing the upper reaches of the ASL and should thus be comparable to results obtained by finely-resolved LES. The results (Fig. \ref{fig:Cabauw_var}) show that, as already observed for anisotropy (see Fig. \ref{fig:Cabauw_anis}), the behaviour of all variance ratios depends on height - not just $R_{if}$. In fact, the upper levels (60 - 180 \,m) point to different flow dynamics than the lowest measurement level (3 \,m), and this difference is examined separately as it hints to a possible role of wall-blocking.

At upper levels (blue colours in Fig. \ref{fig:Cabauw_var}), the behaviour of variance ratios follows the expected patterns suggested by the reduced \textbf{Model R}, albeit with a different magnitude. The contribution of $\overline{u'^2}$ to the total $e$ decreases as $-R_{if}$ increases and the atmosphere becomes more convective (Fig. \ref{fig:Cabauw_var}a). The reduced \textbf{Model R} (Eq. \ref{eq:Reduced} and full line in Fig. \ref{fig:Cabauw_var}) does support this behaviour when looking at the individual 30 min averaging periods at upper heights (dark blue points in Fig. \ref{fig:Cabauw_var}a). The bin averages, however, suggest that the reduced model overestimates this behaviour, as the contribution of $\overline{u'^2}/e$ is neither as large in the neutral regime as the model predicts, nor as small in highly convective regime where $\overline{w'^2}/e$ is predicted to dominate. Instead, a tendency towards an energy equipartition  (all variance ratios equal 2/3) is observed in convective conditions, which means that even at z = 180 \,m, the significance of the wall effects has not waned. Similarly, the wall-normal variance $\overline{w'^2}$ shows increasing contribution to the total $e$ as $-R_{if}$ increases (as anticipated) but at a lower rate, suggesting that as buoyancy generation begins to exceed shear generation ($-R_{if} > 1$), the production term in the vertical direction itself appears thwarted by the wall effect. The spanwise variance ratio $\overline{v'^2}/e$, on the other hand, shows almost no variation with $-R_{if}$ in agreement with having no source of its own \citep[see,][]{BouZeid2018}. The small increase of $\overline{v'^2}/e$ at high $-R_{if}$ corresponds to increasing inability to define a coordinate system in free convective regimes, where convective cells dominate the flow dynamics (see Sect. \ref{sec:discussion}), leading to horizontally isotropic turbulence (streamwise and spanwise variances contribute equally to the total $e$).

These results indicate that even at large distances from the wall, adjustments to the reduced model \textbf{Model R} (Eq. \ref{eq:Reduced}) are necessary. The data suggest both a significantly larger Rotta constant (corresponding to a weaker variation with $R_{if}$), as well as a lower $\overline{u'^2}/e$ and $\overline{w'^2}/e$, and larger $\overline{v'^2}/e$. This adjustment can be achieved with the addition of wall effects in the Rotta model through $a_u, a_v, a_w$ (cf. Eq. \ref{eq:Reduced_anis}) to capture the behaviour of variance ratios (\textbf{Model Ra}, shown with dashed lines in Fig. \ref{fig:Cabauw_var}). This accounting is needed to allow the base anisotropy of the flow caused by wall blocking to be preserved. In this representation, wall blocking causes the $\overline{w'^2}/e$ to receive disproportionately smaller share of energy ($a_w = 0.58$) from the $\overline{u'^2}/e$ at the expense of increasing $\overline{v'^2}/e$ ($a_v = 1.15$) in shear-driven conditions ($-R_{if} < 0.1$). Such an anisotropic model captures the energy anisotropy of the data better than the reduced model (compare full and dashed lines in Fig. \ref{fig:Cabauw_anis}), and confirms that wall effects can persist to large heights \citep[cf.,][]{HuntGraham1978}. In fact, the ratio of measurement height to Eulerian integral length scale for $w'$ remains close to unity even as such unstable conditions are approached (see Sect. \ref{sec:spanwise}). The Rotta model with wall blocking  and adjusted Rotta constant (dashed black line in Fig. \ref{fig:Cabauw_var}) is able to capture the general behaviour of variance ratios at these heights. Having a variable Rotta constant for each component versus fixing it at $c =3$ has a minor influence on the results at these heights, implying that the anisotropy of the energy dissipation that is supposed to be represented by unequal coefficients is not significant.  

At the lowest measurement level (brown colours in Fig. \ref{fig:Cabauw_var}), the behaviour of spanwise $\overline{v'^2}/e$ and the wall-normal $\overline{w'^2}/e$ variance ratio diverges significantly from the predictions of the \textbf{Model R} (Eq. \ref{eq:Reduced}). Stratification appears to have a surprisingly limited effect on $\overline{w'^2}/e$, which changes only modestly with increasing instability, indicating persistent anisotropy at z = 3 \,m, irrespective of stratification. In fact, counter-intuitively, $\overline{w'^2}/e$ decreases with increasing instability, contrary to predictions of the \textbf{Model R}. On the other hand, $\overline{v'^2}/e$ increases from low values in neutral stratification to values exceeding the streamwise variance $\overline{u'^2}/e$ for the majority of the stability range. This increase occurs despite the streamline coordinate system used. The minimum in $\overline{w'^2}/e$ and maximum in $\overline{v'^2}/e$ are found in mildly convective conditions ($-R_{if} \sim 1$), beyond which the lack of data prevents further definitive conclusions. The \textbf{Model Ra} fails in predicting this behaviour as well, and would require a negative Rotta constant $c_w$ in the wall-normal direction to capture a decreasing contribution of $\overline{w'^2}/e$ with increasing instability. On the other hand, no simple modification to the reduced model is able to capture the increase of $\overline{v'^2}/e$. 

The counter-intuitive result of decreasing contribution of $\overline{w'^2}/e$ with increasing instability has been previously observed \citep{Nguyen2013,Ding2018}. It was attributed to a `negative return-to-isotropy' - taking energy from the least energetic component and depositing it into the most energetic, countering the expected flow of energy among components. Our results, however, show that both the most energetic component ($\overline{u'^2}/e$) and the least energetic component ($\overline{w'^2}/e$) appear to lose energy through the return to isotropy, sustaining an increase in the spanwise variance. 

\begin{figure}
    \centering
        \includegraphics[width = 1\linewidth]{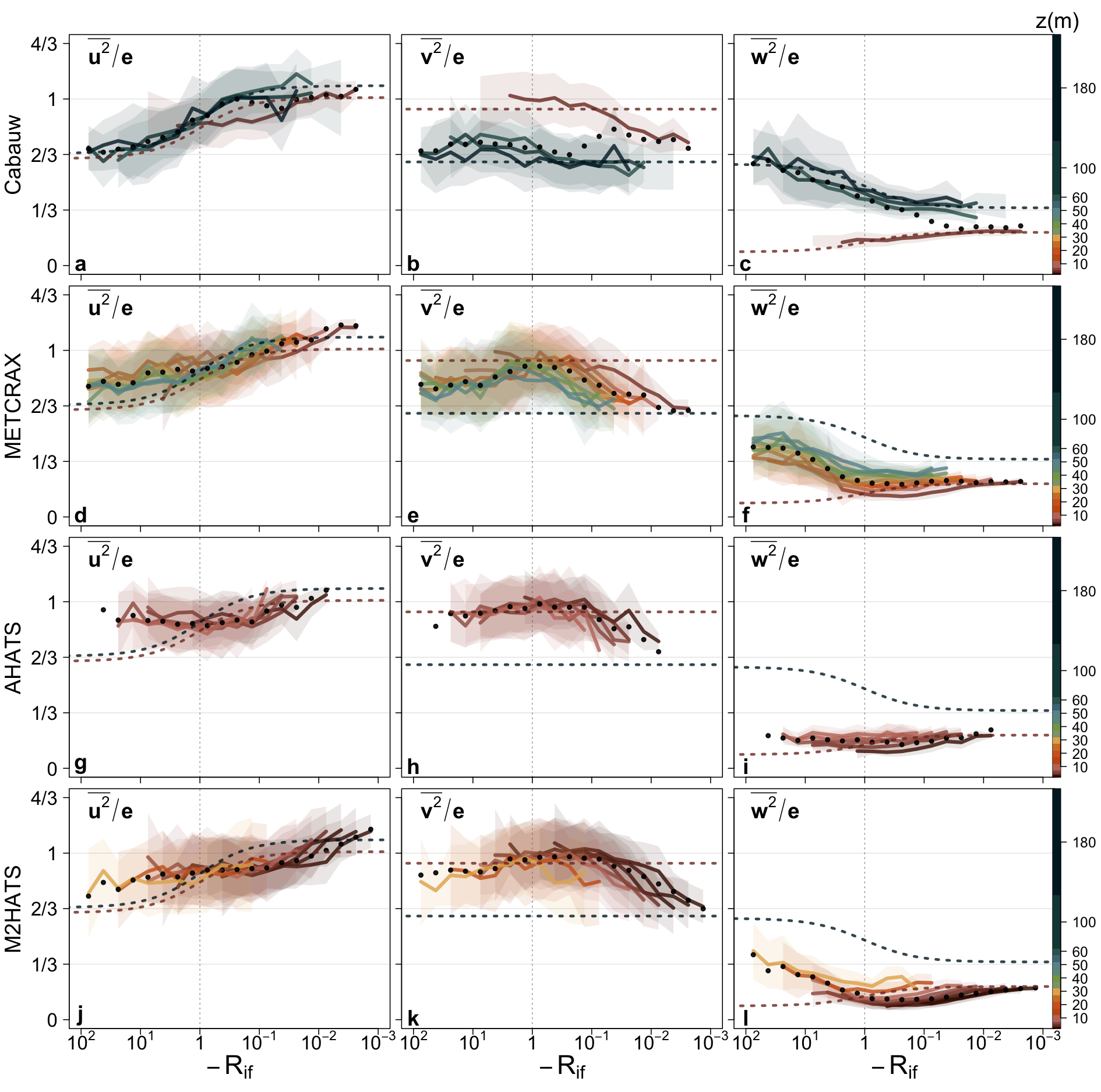}
    \caption{Bin averages of (a,d,g,j) streamwise $\overline{u'^2}/e$, (b,e,h,k) spanwise $\overline{v'^2}/e$ and (c,f,i,l) wall-normal $\overline{w'^2}/e$ velocity variance ratio as a function of $R_{if}$ for Cabauw (a-c), METCRAX (d-f), AHATS (g-i), and M2HATS (j-l) towers. Full coloured lines are logarithmically spaced bin averages for each measurement height (colours), while shading is the interquartile range. Black points are bin averages of all the data, irrespective of height. The dashed curve corresponds to the predictions of \textbf{Model Ra} (Eq. \ref{eq:Reduced_anis}) where the model anisotropy and Rotta constants were obtained from a robust linear fit for the upper levels (60 - 180m, black) and first level (3m, brown) of the Cabauw dataset. Vertical dashed line corresponds to $-R_{if} = 1$}
    \label{fig:Flat_var}
\end{figure}

The decrease of wall-normal velocity variance and the increase of spanwise variance with $-R_{if}$ is a characteristic of other datasets as well (Fig. \ref{fig:Flat_var}). All examined datasets exhibit a clear decrease of $\overline{w'^2}/e$ with increasing instability at heights below z = 10 - 15 \,m, reminiscent of the one observed at Cabauw, although the \textbf{Model Ra} adapted for Cabauw (red dashed line, with a negative Rotta constant) captures only a general behaviour and not the subtleties of each dataset. At the same time, all the other datasets show a pronounced increase of $\overline{v'^2}/e$ with increasing instability, from low values in near-neutral to weakly unstable stratification ($-R_{if} < 0.1$), and a peak around or below $-R_{if} = 1$. This increase occurs through a much deeper layer (up to 50 \,m) than the decrease of $\overline{w'^2}/e$, which is limited to heights close to the ground. Thus, the observations from other datasets show that the minimum in vertical and maximum in spanwise variance are in fact unrelated. 

At high instabilities, the behaviour of variance ratios is influenced by individual site characteristics. Still, the data do suggest that at lower observational heights ($z < 10$m), $\overline{v'^2}/e$ remains high and constant with increasing instability beyond the peak region, and is larger than $\overline{u'^2}/e$. Instead, at higher observational heights (z = [10 - 50 m]), turbulence appears to be more horizontally isotropic as $\overline{u'^2}/e$ and $\overline{v'^2}/e$ have similar values. 
These observed characteristics of velocity variance ratios remain visible when data are grouped as a function of the local stability parameter $\zeta$ instead of $R_{if}$ (cf. Fig. \ref{fig:KaderYaglom}), and, therefore, the trends are not an artifact of the presentation against the flux Richardson number (cf. Fig. \ref{fig:KaderYaglom}).

\subsection{Extended model of the Reynolds stresses}
\label{sec:extended}

As seen in Sect. \ref{sec:simple}, the reduced model \textbf{Model Ra} with wall-blocking and larger Rotta constant ($c=3$) reproduces the variance ratios at upper levels ($z > $ 60 m) of the Cabauw dataset reasonably well. At lower levels ($z = $3 m), this model is unable to capture the observed behaviour of $\overline{w'^2}/e$ or $\overline{v'^2}/e$. The extended \textbf{Model E} (Eqs. \ref{eq:FullX} - \ref{eq:FullZ}) is now used to assess if the neglected (transport and rapid pressure) terms can account for these differences.

\begin{figure}
    \centering
        \includegraphics[width = 1\linewidth]{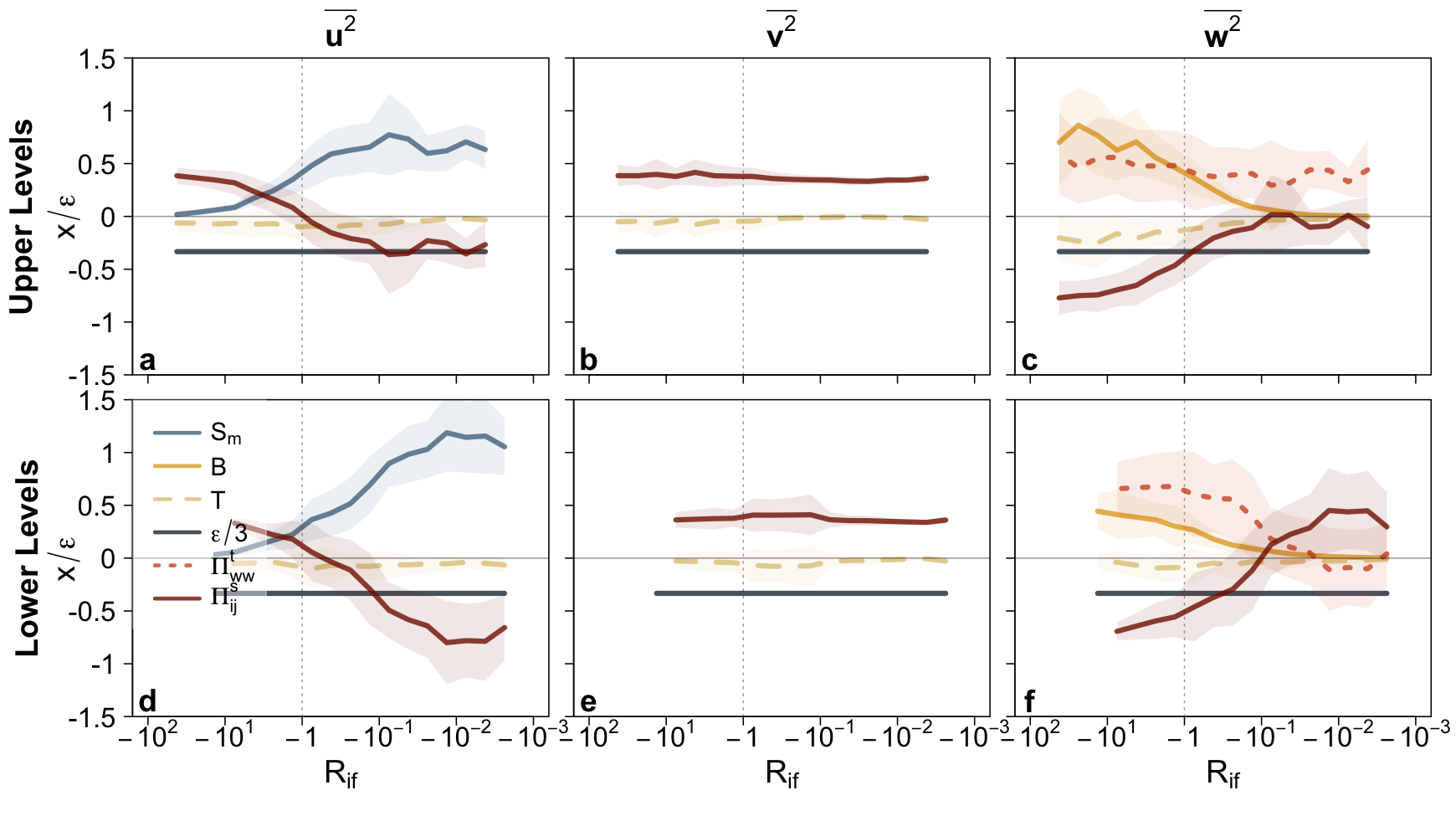}
    \caption{Reynolds stress budget terms (colour) normalized by the total dissipation ($\varepsilon$) for the (a,d) streamwise, (b,e) spanwise, (c,f) wall-normal variance, as functions of the flux Richardson number for the Cabauw dataset. The Reynolds stress budget terms are shown for 60m - 100m levels (a - c) the 3m level (d - f). The names of the budget terms are defined in Eq. \ref{eq:Reynolds}.}
    \label{fig:ReynoldsStresses}
\end{figure}

The behaviour of Reynolds stress budget terms as a function of $R_{if}$ (Fig. \ref{fig:ReynoldsStresses}) highlights that the presence of the pressure transport, already noted to be an important term in the TKE budget, is the dominant term driving the shape of the pressure-strain contributions in the wall-normal variance budget at lower levels (note the opposite behaviour of pressure transport and pressure-strain terms in Fig. \ref{fig:ReynoldsStresses}d-f), and that the wall-normal component starts to lose energy through the pressure redistribution processes already at $-R_{if} = 0.1$ \citep[the simplified model predicts this switch to occur at  $-R_{if} = 0.5$, ][]{BouZeid2018}, to the advantage of a slightly growing spanwise variance. At the same time, the pressure-redistribution for the streamwise variance budget becomes positive only at $-R_{if} = 1$ \citep[the simplified model predicts this switch to occur at $-R_{if} = 2$, ][]{BouZeid2018}. Since the slow return to isotropy through the Rotta term is itself proportional to the variance ratios, which, as we saw, behave non-linearly, we can expect the rapid terms to account for this non-linearity and therefore play an important role in driving the pressure-strain process. 

For the upper levels of the Cabauw dataset, \textbf{Model Ra} was already able to capture the general characteristics of the observed variance ratios. The additional terms that form the \textbf{Model E} are therefore expected to have a limited effect on the results, apart from modifying the value of the Rotta constant. This is confirmed by the Reynolds stress budgets themselves, which show that the pressure-strain is dominated by the buoyancy and shear production terms (Fig. \ref{fig:ReynoldsStresses}a-c). The inclusion of the rapid terms would, in fact, allow the Rotta constant to attain lower values (making them closer to accepted values from laboratory studies). 

\begin{figure}
    \centering
        \includegraphics[width = 1\linewidth]{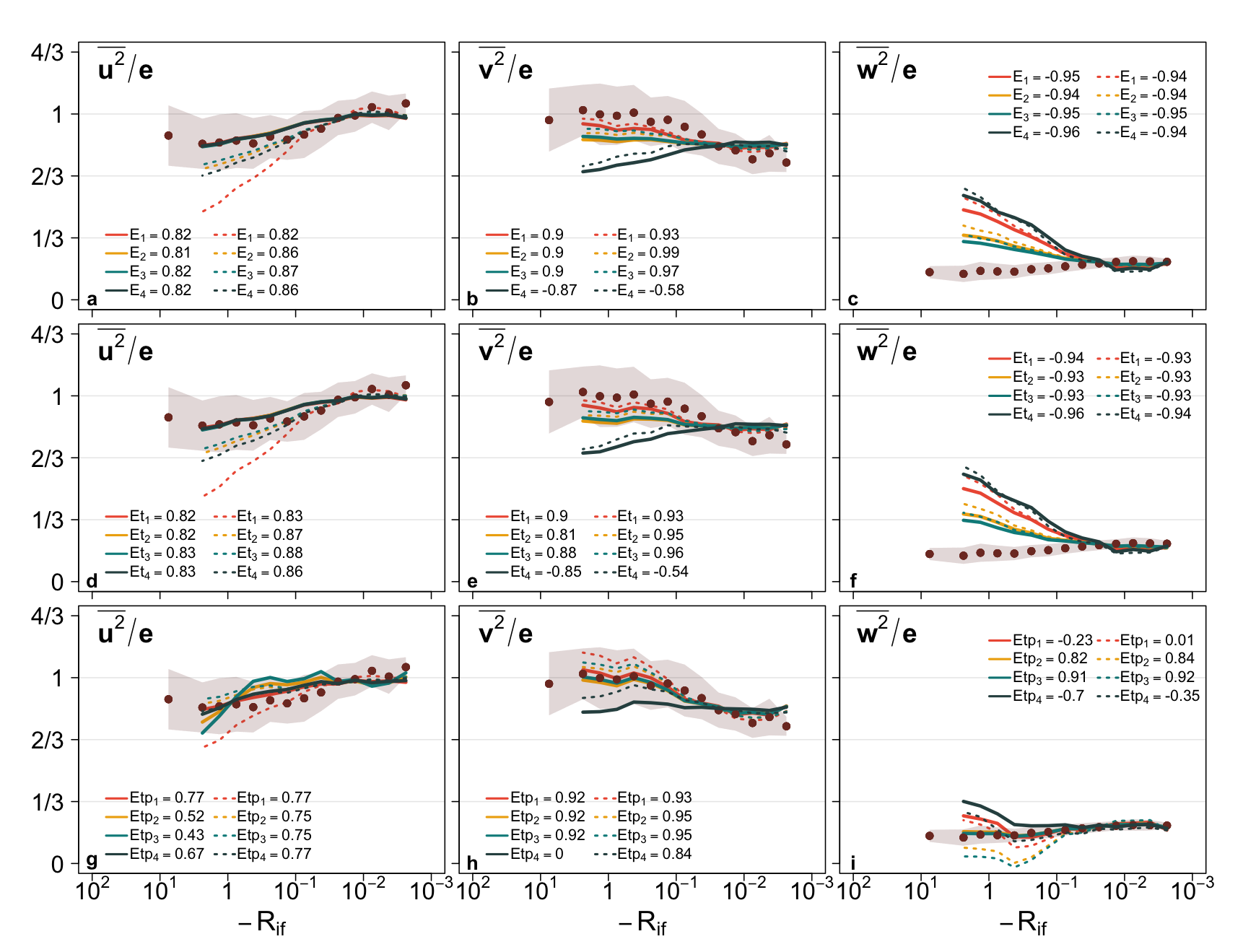}
    \caption{Predictions of the velocity variance ratios as a function of $R_{if}$ for the lowest measurement level of the Cabauw tower, for the the \textbf{Model E}. The model has (a-c) no transport terms (\textbf{Model E}) , (d-f) turbulent transport included (\textbf{Model Et}), (g-i) both turbulent and pressure transport included (\textbf{Model Etp}). The thick lines correspond to models with variable Rotta constant and wall blocking \textbf{Model E}, while thin dashed lines correspond to the non-linear Rotta \textbf{Model E$_n$}. The different combinations of rapid terms are shown in colour: \textbf{Model E$_1$} with no rapid terms (red), \textbf{Model E$_2$} with $C^u_{B} =0.3$ and $C^u_{S1} = C^u_{S2} = 0.6$ (yellow), \textbf{Model E$_3$} with $C^u_{B} =0.6, C^u_{S1} = C^u_{S2} = 0.6$ (turquoise), and \textbf{Model E$_4$} $C^u_{B} = 0.6, C^u_{S1} = 12/7, C^u_{S2} = 0$. Numbers in the legend refer to the correlation coefficient between the binned observed data and binned model data. }
    \label{fig:Cabauw_Model}
\end{figure}

The ultimate test of the model, however, is its ability to reproduce the variance ratios at low levels, where, apart from an increase of $\overline{v'^2}/e$, the decrease of $\overline{w'^2}/e$ is also observed (Fig. \ref{fig:Cabauw_Model}). We thus test the importance of different components of \textbf{Model E}: without and with turbulence transport \textbf{Model Et} and pressure transport \textbf{Model Etp}, and addition or not of rapid pressure-strain terms with varying constants found in literature (different numbers in the subscript, see table \ref{tab:tableModels}), including using the non-linear Rotta model as an alternative slow pressure-strain parameterization (subscript $_n$). 

The results show that the streamwise variance ratio $\overline{u'^2/e}$ (Fig. \ref{fig:Cabauw_Model}a,d,g) is reasonably captured by all the models, whether they include turbulent and pressure transport or not. However, both the wall blocking and Rotta constant require adjustments. If the non-linear Rotta model is used, then only the model that includes both transport terms, as well as the rapid isotropization of production terms (\textbf{Model Etp$_n$}) is able to describe the observed behaviour. 

The outcomes are different if the spanwise variance ratio $\overline{v'^2/e}$ is examined though (Fig. \ref{fig:Cabauw_Model}b,e,h), as only a subset of model versions are able to account for the increase of spanwise variance with increasing instability. The first important result is the recognition that the production and dissipation are not balanced, thus even the model that does not include transport or rapid terms \textbf{Model E$_1$} outperforms \textbf{Model Ra}. The second is that the inclusion of rapid terms without the inclusion of pressure transport \textbf{Model E$_{2,4}$} actually deteriorates the model performance. If the pressure terms are included, though, then all but the last version of the model  with the rapid terms (\textbf{Model Etp$_4$}) reproduce the observed behaviour of this velocity variance ratio. While for Cabauw, the pressure transport does not appear to be crucial in driving the increase of spanwise variance, in other datasets, however, this increase cannot be reproduced unless both turbulent and pressure transport are added (cf. Figs. \ref{fig:Metcrax_Model} and \ref{fig:M2HATS_Model} in Appendix \ref{appendixOtherFigures}). 

Finally, only one version of the model is able to describe the observed decrease of the wall-normal velocity variance $\overline{w'2}/e$ (Fig. \ref{fig:Cabauw_Model}c,f,i), and that is the model that includes both the turbulent and pressure transport, and the rapid terms (\textbf{Model Etp$_{2,3}$}). Actually, all the models that do not include the pressure transport would require a negative Rotta constant to match the observations (see the negative value of correlation coefficients in the legends of Fig. \ref{fig:Cabauw_Model}). In addition, the non-linear Rotta model (Eq. \ref{eq:Rotta_nonlin}) would still require adjustments in order to capture the observed behaviour of the variance ratios.

The main outcomes of this analysis are that the processes that determine the behaviour of near-surface anisotropy are tightly coupled to transport processes, including both the turbulent and pressure transport. No existing reduced model captures these processes. 
In addition, irrespective of the transport terms, the combination of rapid terms that does not include the vorticity term (\textbf{Model E$_4$}) is unable to represent the observed behaviour of either the spanwise or the wall-normal variance, pointing to the importance of vorticity in the processes that cause deviations from the behaviour expected from the reduced models. Equivalent results are obtained for other datasets (see Figs. \ref{fig:Metcrax_Model} and \ref{fig:M2HATS_Model} in Appendix \ref{appendixOtherFigures}).

\subsection{Wall-normal variance minima and spanwise variance maxima}
\label{sec:spanwise}

The importance of the pressure transport and rapid pressure-strain terms in explaining the observed behaviour of spanwise and wall-normal variances poses the question of their origin. A possible source could be turbulence organization into coherent structures, the nature of which changes with changing stratification \citep[e.g., ][]{Li2011,Salesky2017,Zilitinkevich2021,LiHatchinsMarusic2022}. \cite{Salesky2017} have shown that coherent structures in a convective boundary layer undergo a transition from convective rolls to convective cells at around $-z_i/L = 15 - 20$, where $z_i$ is the mixed layer height and $L$ is the Obukhov length based on surface fluxes. \cite{Salesky2018} and \cite{LiHatchinsMarusic2022} have shown that at a stability parameter $-\zeta \sim 1$, the wall-attached coherent structures change their inclination angles, and that for higher instabilities their aspect ratios undergo a transition, with the size of coherent structures increasing both in the wall-normal and spanwise directions, relative to their streamwise extent.
The question is therefore what is the influence of pressure transport in this process \citep[cf.][]{Cuxart2002}, and whether the streamline curvature associated with this changing nature of coherent structures and their inclination angles can cause turbulence to undergo rapid distortion. Recently, \cite{Mosso2025} have shown that the degree of anisotropy is coupled with the rapid distortion parameter. 

\begin{figure}
    \centering
        \includegraphics[width = 1\linewidth]{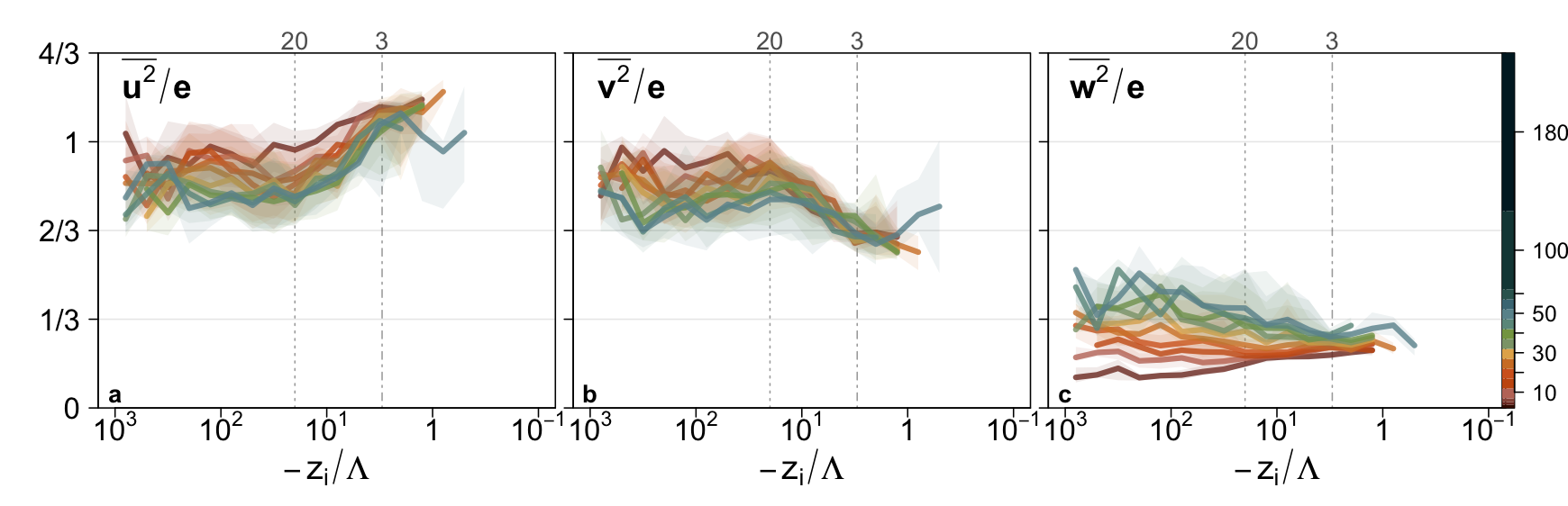}
    \caption{Velocity variance ratios of a) $\overline{u'^2}/e$, b) $\overline{v'^2}/e$, and c) $\overline{w'^2}/e$ as function of $z_i/\Lambda$ for the METCRAX II dataset. Here, $z_i$ is the PBL height obtained from ERA-Land reanalysis, and $\Lambda$ is the local Obukhov length. Full lines correspond to averages over logarithmically spaced bins of $z_i/\Lambda$ for different measurement heights (colours), while the shading is the interquartile range. Vertical dash-dotted (dashed) line corresponds to $-z_i/\Lambda = 3$ ($-z_i/\Lambda = 20$) respectively.}
    \label{fig:MetCrax_ZiLambda}
\end{figure}

To explore if the observed minimum in $\overline{w'^2}/e$ and increase or peak of $\overline{v'^2}/e$ are indeed the result of the organization of turbulence into different types of coherent structures, we focus on the METCRAX II experiment, as its high vertical resolution and large heights allow testing of this hypothesis. Figure \ref{fig:MetCrax_ZiLambda} shows that if the variance ratios are plotted as a function of $-z_i/\Lambda$ (where $\Lambda$ is the local Obukhov length) instead of the local flux Richardson number, there is a clear change of behaviour in velocity variance ratios at $-z_i/\Lambda = 20$ consistent with LES findings by \cite{Salesky2017}. In fact, Fig. \ref{fig:MetCrax_ZiLambda} suggests the existence of three regimes. For small $-z_i/\Lambda < 3$ corresponding to near-neutral stratification and organization of turbulence into hairpins and streaks associated with wall-attached eddies \citep{Hutchins2012}, the behaviour of variance ratios is independent from stratification (ratios are constant), streamwise variance $\overline{u'^2}/e$ dominates the TKE, while spanwise $\overline{v'^2}/e$ and wall-normal $\overline{w'^2}/e$ variances are small. In an intermediate stratification range $-z_i/\Lambda = [3 - 20]$, corresponding to predominance of convective rolls, a large increase of spanwise variance and a concomitant decrease of streamwise variance occur with no effect on the wall-normal variance. 
Finally, under very unstable stratification 
$-z_i/\Lambda > 20$ where turbulence organization is dominated by convective cells, streamwise and spanwise variance ratios are horizontally isotropic, while the wall-normal variance finally starts to exhibit a pronounced change with height: a clear decrease at low levels ($z < $15 m) and a clear increase at higher levels ($z > $30 m). It may be conjectured that this low-level behaviour corresponds to the wall-blocking (pressure echo) of large convective structures. In fact, a physical rationale for this low level of vertical velocity near the ground can be conceptualized, given the generation of warm plumes and parcels in contact with the hot surface. These hot parcels will have low vertical kinetic energy, but their potential energy will be very high (warmer than the surroundings) and, as they rise, they are able to accelerate (conversion of potential to kinetic energy via buoyancy generation) and increase the vertical variance at higher levels.

\begin{figure}
    \centering
        \includegraphics[width = 1\linewidth]{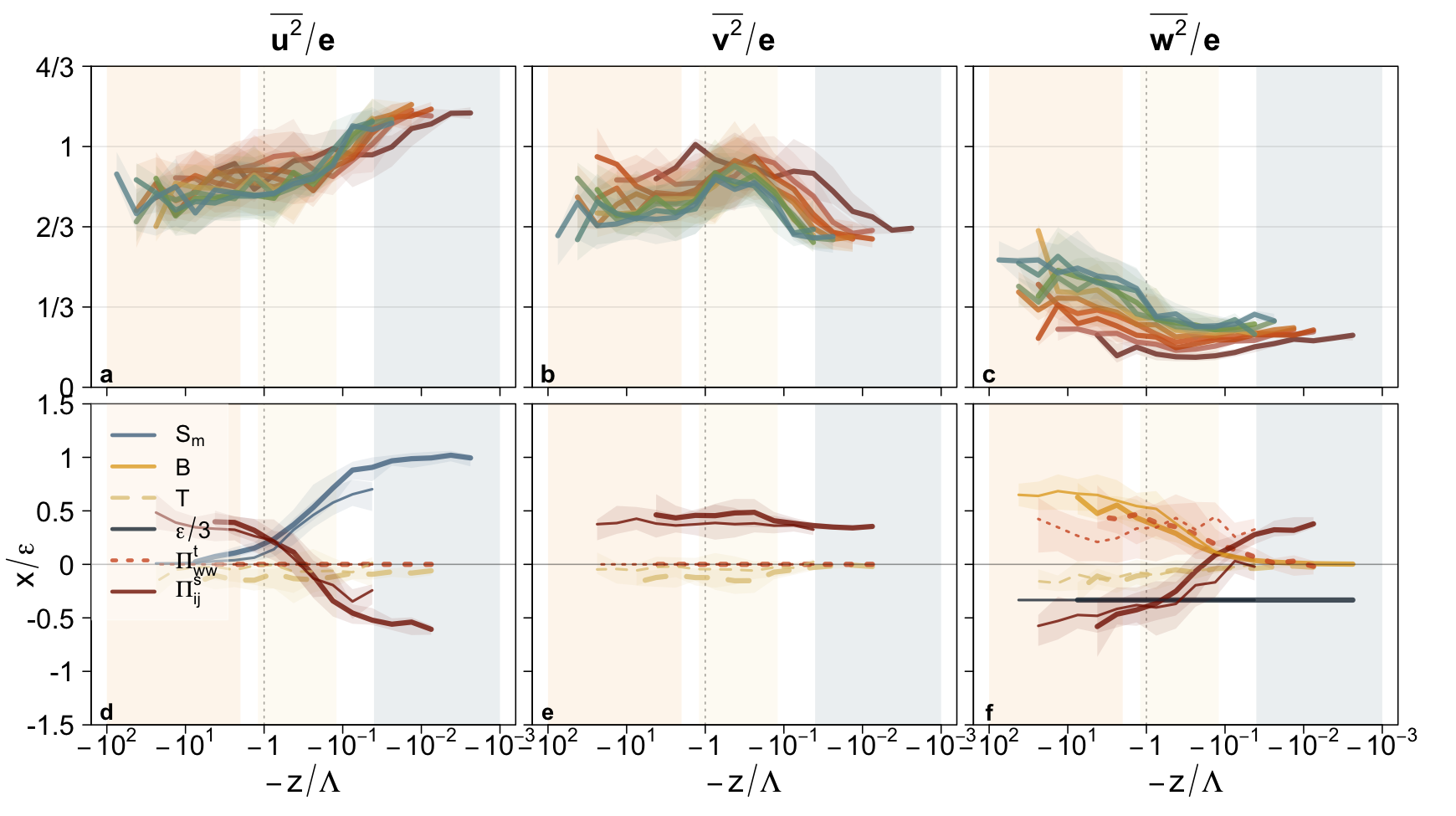}
    \caption{a - c) Velocity variance ratios and (d - f) terms of the Reynolds stress budgets normalized by the dissipation rate as a function of the local stability parameter ($z/\Lambda$) for a,d) streamwise, b,e) spanwise and c,f) wall-normal variance for the METCRAX II dataset. Here, the variable names are defined in Eq. \ref{eq:Reynolds}. The full lines and shading in (a-c) are bin averages and interquartile ranges for each height (colors), while in (d-f) the thick lines correspond to medians over $z=$3 - 10 m, and thin lines medians over $z=$35 - 45 m. Shaded areas correspond to dynamic ($0<-\zeta<0.04$, blue), dynamic-convective ($0.12<-\zeta<1.2$, yellow), and convective ($-\zeta>2$, orange) sub-ranges of \cite{KaderYaglom1990}.}
    \label{fig:KaderYaglom}
\end{figure}

These three regimes can be shown to be related to the three sub-layers of \cite{KaderYaglom1990}, governed by different scaling parameters due to their differing dynamics (Fig. \ref{fig:KaderYaglom}). The neutral stratification, characterised by turbulence streaks and dominance of streamwise variance, corresponds to the dynamic sub-layer ($0<-\zeta<0.04$). This is the regime where turbulent ($T$) and pressure ($\Pi_{ww}^t$) transport terms are generally insignificant in all of the variances. Moreover, the pressure-strain is constant and its streamwise component ($\Pi_{uu}$) is more negative than $\varepsilon/3$ (Fig. \ref{fig:KaderYaglom}d-f). At the transition to the dynamic-convective sub-layer ($0.12<-\zeta<1.2$), a rapid increase of the importance of transport terms and an increase of ($\Pi_{uu}^t$) occurs. These increases are accompanied by a decrease in ($\Pi_{ww}^t$) that both cross over $\varepsilon/3$ at the onset of the dynamic-convective sub-layer. The spanwise variance is now clearly gaining more energy from both the streamwise and wall-normal components, and conveying it through the turbulent transport term. The convective sub-layer ($-\zeta>2$) is not observed at the lowest measurement height. At higher wall-normal distances, however, the results suggest that the Reynolds stress budget terms again become more or less constant with increasing stratification. These findings from the anisotropy analysis here are consistent with the premise of directional dimensional analysis \citep{KaderYaglom1990}, and suggest a possible connection between the different ranges and transitions in the structure of turbulence organization. 

Finally, the additional confirmation that the organization of turbulence into coherent structures and wall blocking plays a dominant role in the behaviour of variance ratios with increasing instability comes from inspecting the following measures: 
\begin{enumerate}
    \item The ratio of Eulerian integral length scale of wall-normal variance normalized by height ($\lambda_w/z$) provides information on the importance of wall blocking.
    \item The ratio of Eulerian integral length scale of the horizontal velocity variance ($\lambda_u$) normalized by the shear length scale ($L_s={\sqrt{\overline{u'^2}}}/|dU/dz|$) provides information on whether shear acts to limit the largest turbulent scales in the streamwise direction \citep{Jacobitz1999} by `shredding' eddies larger than the shear length scale
    \item The non-dimensional skewness of wall-normal velocity ($\overline{w'^3}/\overline{w'^2}^{3/2}$) indicates if convective structures (strong updrafts limited in space with large areas of weak downdrafts) are found in the flow. 
    \item The non-dimensional time scale associated with rapid distortion ($\tau_{\epsilon}u_*/\kappa z$) \citep[][chapter 11.4]{Pope2000}, provides  information on the degree of non-equilibrium of the flow with respect to its forcing, and therefore the importance of rapid distortion. It represents the ratio of mean shear time scale to memory time scale ($\tau_{\epsilon}$), where we have adapted the mean shear time scale to its logarithmic value ($\kappa z/u_*$). 
\end{enumerate}

\begin{figure}
    \centering
        \includegraphics[width = 0.8\linewidth]{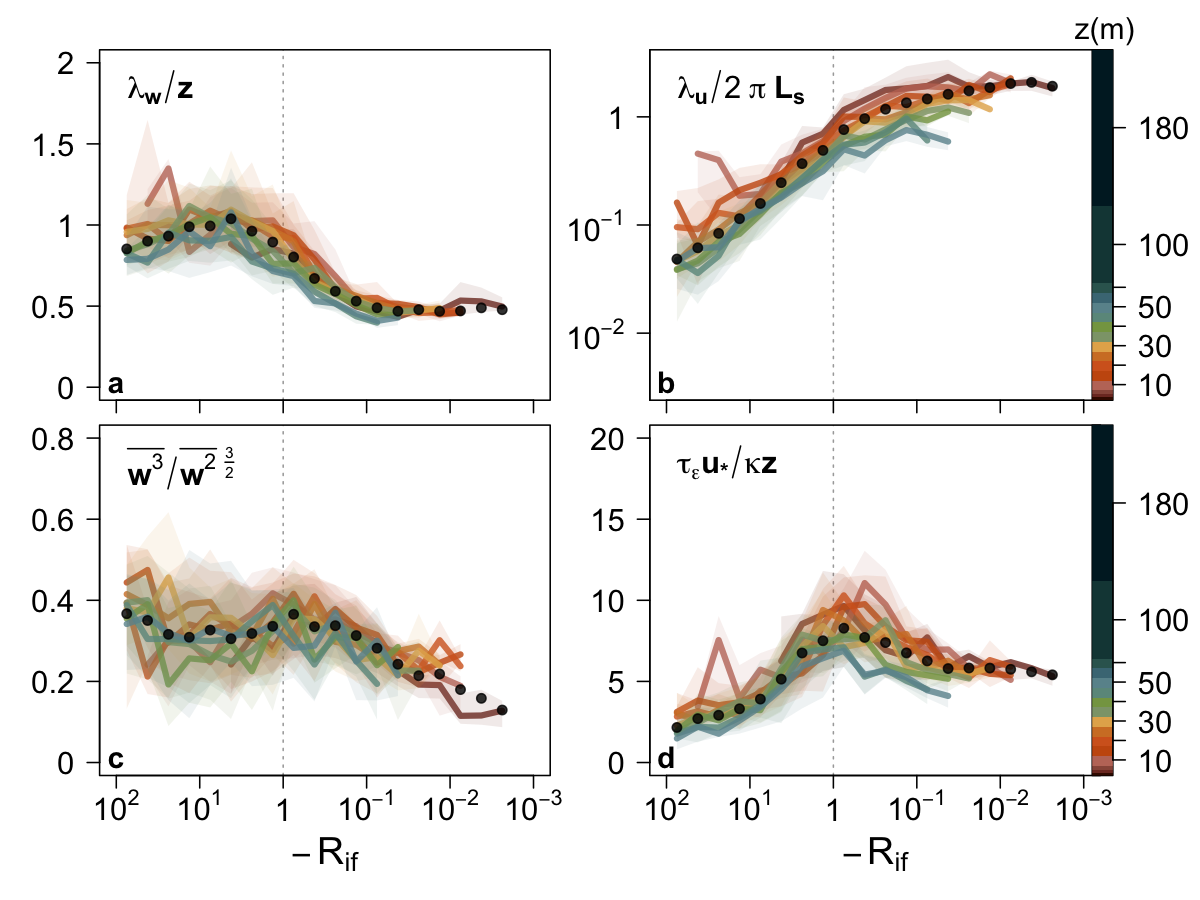}
    \caption{a) Eulerian integral length scale of wall-normal velocity normalized by the observational height ($\lambda_w/z$), b) Eurlerian integral length scale of streamwise variance normalized by the shear length scale ($\lambda_u/L_s$), c) non-dimensional skewness of wall-normal velocity ($\overline{w'^3}/\overline{w'^2}^{3/2}$), and d) a non-dimensional rapid distortion timescale ($\tau_{\epsilon}u_*/\kappa z$), as function of $R_{if}$. Full lines correspond to bin averages of different measurement heights (colours), while the shading is the interquartile range. Vertical dashed line corresponds to $-R_{if} = 1$.}
    \label{fig:MetCrax_RDT}
\end{figure}

All four measures have a consistent behaviour (Fig. \ref{fig:MetCrax_RDT}) and point to sources of anisotropy in the three regimes, as well as the importance of regime transition at $-R_{if} \sim 1$. In the near-neutral regime ($-R_{if} < 0.1$) dominated by horizontal structures, integral length scale is proportional to height ($\lambda_w/z \simeq 0.5$) and turbulence appears free from wall-blocking effects at all heights ($\lambda_w/z < 1$) (Fig. \ref{fig:MetCrax_RDT}a). The shear length scale acts as the limiting length scale ($\lambda_u/L_s > 1$) thwarting the energy in the streamwise direction and allowing it to accumulate at large scales of the spanwise variance $\overline{v'^2}$, for which the corresponding shear scale is very large owing to $|dV/dz|\approx 0$ (Fig. \ref{fig:MetCrax_RDT}b). We observe this through the increase of spanwise variance in this range (cf. Fig. \ref{fig:Flat_var}).
At the same time, skewness is low, and the rapid distortion timescale ($\tau_{\epsilon}u_*/\kappa z \sim 3 - 6$) attains values typical of mean shear flows (Fig. \ref{fig:MetCrax_RDT}c,d). Here, the small changes of stratification have little effect on the flow characteristics, as already observed through other measures. Thus in this regime, the increase of spanwise variance is hypothesized to come at the expense of the streamwise variance through the action of shear. 

In the weakly unstable regime ($-R_{if} = [0.1 -- 2.5]$), however, the flow experiences simultaneous onset of importance of wall blocking ($\lambda_w/z$ approaching one), and onset of decreasing importance of shear ($\lambda_u/L_s < 1$), and an increase of non-dimensional skewness ($\overline{w'^3}/\overline{w'^2}^{3/2}$) indicating growing importance of convective updrafts in the wall-normal velocity statistics associated with the increasing inclination angles \citep{Salesky2018} and the lifting of horizontal structures \citep{LiHatchinsMarusic2022}, even at these low heights within the ASL. We also observe the increasing importance of rapid distortion ($\tau_{\epsilon}u_*/\kappa z$ has a maximum at measurement levels below $z \sim $ 20 m) that suggests that turbulence is out of equilibrium with its forcing in this regime.
In fact, both the skewness and the rapid distortion timescale peak at $-R_{if} \sim 1$, collocated with the peak inclination angle according to \cite{Salesky2018}, and the observed maximum in $\overline{v'^2}/e$ (Fig. \ref{fig:Flat_var}h) and minimum in $\overline{w'^2}/e$ (Fig. \ref{fig:Flat_var}i) due to the dual effect of vertical lifting of coherent structures and rapid distortion effects. In this regime, the flow attains the maximum energy anisotropy close to the surface (cf. Fig. \ref{fig:Cabauw_anis}), and here the spanwise variance receives energy from both the streamwise direction (through the action of shear) and wall-normal direction (through the action of pressure transport).

Finally, in strongly unstable, convective regimes ($-R_{if} > 2$), the importance of wall blocking increases ($\lambda_w/z = 1$) even at 50 m height, both shear and the attendent the rapid distortion as a mechanism loses importance, while the convective surface layer remains dominated by convective updrafts, but to a lesser degree. This same behaviour is consistently observed at all the sites (see Fig. \ref{fig:Flat_RDT} in the Appendix \ref{appendixAnisotropicDissipation}). 

\subsection{Scalewise results}
\label{sec:scalewise}

\begin{figure}
    \centering
        \includegraphics[width = 1\linewidth]{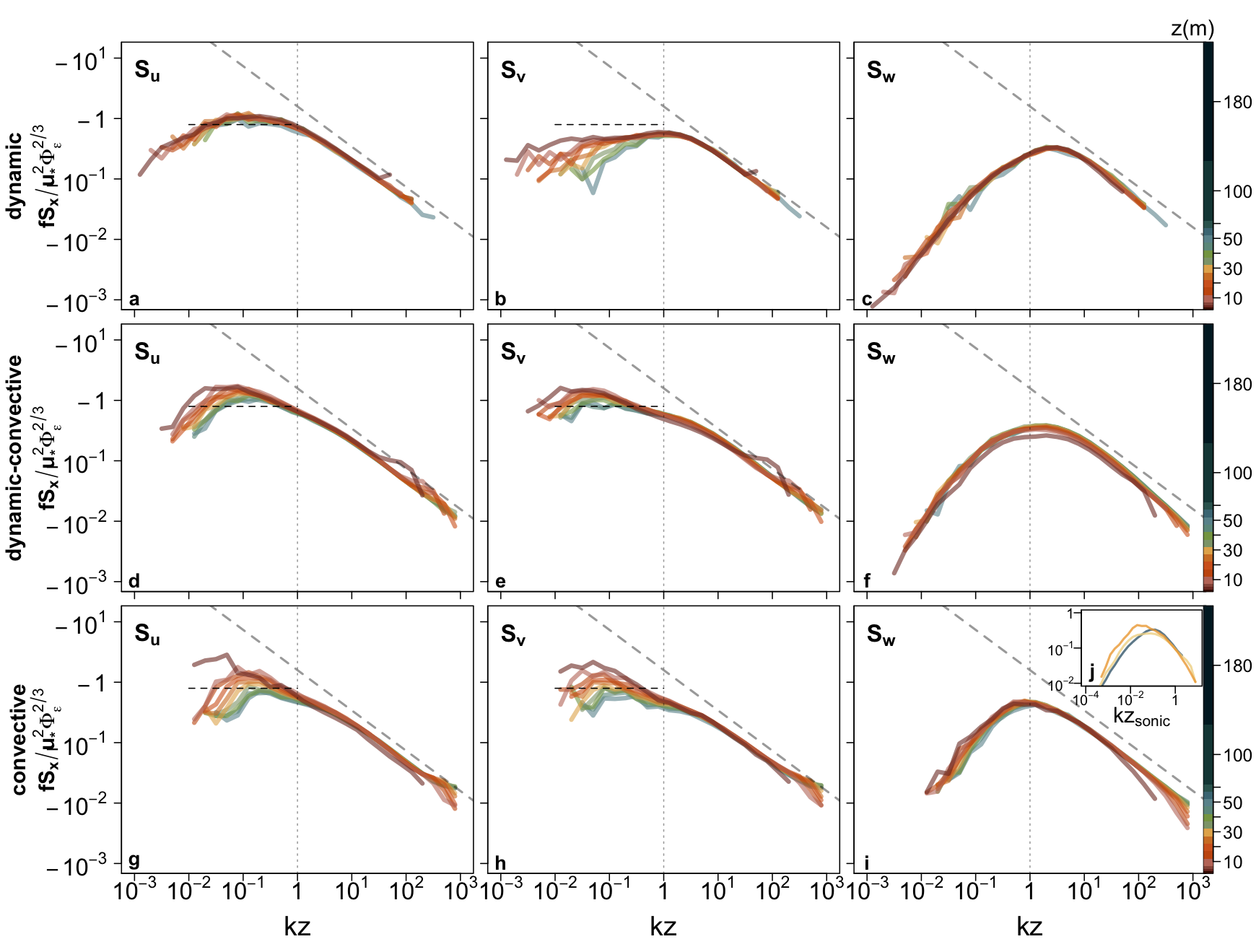}
    \caption{Scaled spectra of (a,d,g) streamwise, (b,e,h) spanwise and (c,f,i) wall-normal velocity, as function of scaled wavenumber $kz$ for METCRAX II dataset, for dynamic ($-\zeta < 0.04$, first row), dynamic-convective ($-\zeta = [0.12 - 1.2]$, middle row), and convective ($-\zeta > 2$) regime. Here, $k = 2\pi f/\overline{U}$ is the streamwise wavenumber obtained from the time domain using Taylor's frozen turbulence hypothesis. 
    Full lines correspond to bin averages over logarithmically spaced $z_i/\Lambda$ for different measurement heights (colours). Diagonal dashed lines indicate a $k^{-5/3}$ slope, while horizontal dashed lines depict $k^{-1}$. Vertical dotted lines indicate $kz = 1$. Insert (j) shows the wall-normal velocity spectra as a function of wavenumber normalized by the sonic path length ($z_{sonic}$) for the lowest measurement level (here 3 m) of the dynamic (blue), dynamic-convective (yellow), and convective (orange) regimes. The inset suggests that the resolved scales in $w'$ far exceed the anemometer averaging path length and variance alterations due to changes in $R_{if}$ cannot be attributed to instrument path averaging. }
    \label{fig:MetCrax_Spectra}
\end{figure}

As a bridge between the aforementioned findings and scalewise turbulence energetics, spectral analysis in the three ASL regimes is considered (Fig. \ref{fig:MetCrax_Spectra}). The scaled spectral densities show that in the near-neutral regime, energy-containing turbulent eddies are attached to the wall, as both the streamwise and spanwise spectra show a $k^{-1}$ slope, where $k$ is the longitudinal wavenumber. The links between the $k^{-1}$ spectral scaling and the attached eddy model at very high Reynolds number have been established elsewhere from a spectral budget and are not repeated here \citep{banerjee2013logarithmic,qin2025asymptotic}. The region with $k^{-1}$ slope is particularly pronounced for the spanwise spectra at low levels and spans more than 2 decades ($kz = [0.01 - 1]$). The spectra also collapse on top of each other for neutral stratification, as expected \citep{kader1991spectra,katul1998theoretical,katul2012existence,huang2022profiles}. At almost all heights, the peak in the spanwise spectrum occurs at lower wavenumbers than in streamwise spectrum (Fig. \ref{fig:MetCrax_Spectra}a,b), which shows a drop in variance at largest scales. This behaviour thus strongly supports the hypothesis proposed in the previous section (see Sect \ref{sec:spanwise}) that the shear (through its length scale $L_s$) limits the energy input into the streamwise variance, thus feeding the spanwise variance.

The spectra in the weakly and strongly unstable regime show marked differences. In weakly unstable, dynamic-convective regime ($-\zeta = [0.12 - 1.2]$) where convective rolls dominate, the spectral energy shows a marked increase at low wavenumbers. Turbulent eddies can still be considered attached to the wall close to the surface (brown lines) as indicated by a small region with a $k^{-1}$ ($kz = [0.03, 0.1] $) in both the streamwise and spanwise spectra. At larger heights, however, the contribution of streamwise variance is decidedly lower than at lower heights. At the same time, shear continues to limit the energy content of the streamwise variance at the expense of the spanwise variance which shows a peak in the spectra at consistently lower wavenumbers than spanwise variance (Fig. \ref{fig:MetCrax_Spectra}d,e).

Finally, in convective stratification ($-\zeta > 2$) dominated by convective cells, both the streamwise and spanwise spectra show a more pronounced height-dependent peak, found at around $kz = 0.2$ and therefore at higher wavenumbers than in weakly unstable stratification. In this region, streamwise and spanwise spectra both show matching area under the spectral curves and the location of the spectral peaks. This behaviour is expected as convective cells are horizontally isotropic both in spectral density and in the size of the dominant eddies.

\begin{figure}
    \centering
        \includegraphics[width = 1\linewidth]{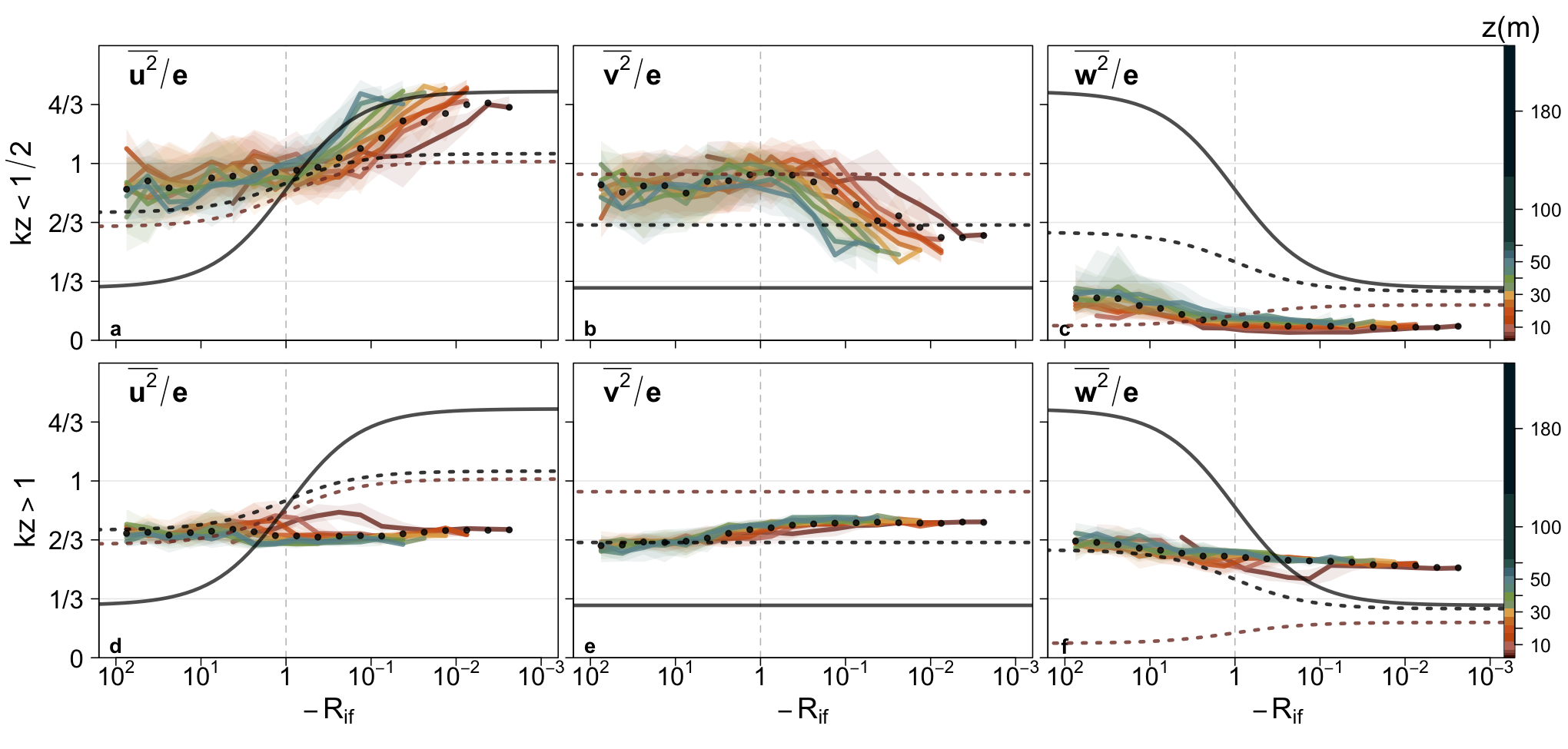}
    \caption{a,d) Streamwise $\overline{u'^2}/e$, b,e) spanwise $\overline{v'^2}/e$ and c,f) wall-normal $\overline{w'^2}/e$ velocity variance ratio as a function of $R_{if}$ for attached eddies assumed as those for which $kz < 1/2$ (upper row) and detached eddies assumed as those for which $kz > 1$ (lower row), as a function of height (colours) for the METCRAX II experiment. Full curve corresponds to the predictions of reduced \textbf{Model R} Eq. \ref{eq:Reduced} with $c = 0.9$, while the dashed curves correspond to the predictions of the reduced model with wall-blocking added \textbf{Model Ra} for the Cabauw dataset (see Fig. \ref{fig:Cabauw_var}) where the Rotta constant has been adjusted to $c=3$. Here, the wall-blocking constants and Rotta constants were obtained from a robust linear fit for the upper levels (60 - 180m, black) and the first level (3m, brown) separately.}
    \label{fig:MetCrax_attached}
\end{figure}

The distribution of energy carried in the attached ($kz < 1/2$) energy-containing eddies, and the detached ($kz > 1$) inertial-subrange eddies, is also a function of stability (Fig. \ref{fig:MetCrax_attached}). The energy content of the attached eddies is closest to the original prediction of the \textbf{Model R}, although due to wall-blocking effects, the vertical variance contribution is still smaller than predicted. Additionally, the rise of the spanwise variance at intermediate flux Richardson numbers occurs already at the largest scales of attached eddies. On the other hand, the inertial-subrange detached eddies also show persistent anisotropy at neutral stratification ($\overline{w'^2}/e < \overline{u'^2}/e, \overline{v'^2}/e$), with the largest energy content of spanwise variance. With the increase of instability, however, the energy content of inertial subrange eddies does show an almost linear tendency towards isotropy in energy distribution, achieved at large Richardson numbers ($-R_{if} >> 1$).
Therefore, the reduced model \textbf{Model R} requires modifications close to the surface, even if only the resolved, energy-containing motions are taken into account.

\section{Discussion}
\label{sec:discussion}

\subsection{Effect of the coordinate system choice on spanwise variance}

Although the previous sections have consistently pointed to the role of pressure transport and action of shear as the sources of spanwise variance, its large contribution to the total TKE deserves a thorough assessment of possible alternative sources, specifically those connected with the coordinate system choice and the role of the Coriolis force. 

\begin{figure}
    \centering
        \includegraphics[width = 0.8\linewidth]{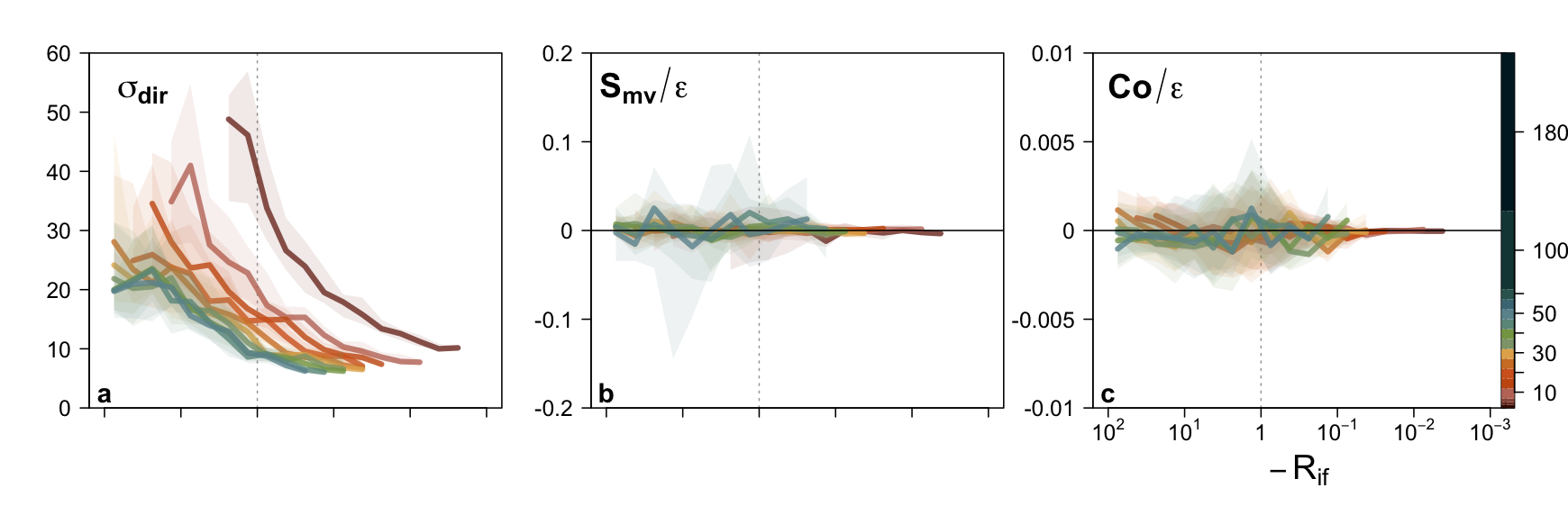}
    \caption{a) Standard deviation of wind direction $\sigma_{dir}$ within a 30-min period, b) spanwise shear production term divided by dissipation $S_{mv}/\varepsilon$ in a fixed coordinate system, and c) Coriolis term in the Reynolds stress budgets divided by dissipation $Co/\varepsilon$, as a function of $R_{if}$ for the METCRAX II experiment. Full lines represent bin averages, and shading the interquartile range, as a function of height (colours).}
    \label{fig:MetCrax_rotation}
\end{figure}

Our analysis rest on the use of double rotation in interpreting the turbulence measurements. This method relies on the fact that a coordinate system can be well defined (i.e., the wind direction does not vary appreciably within the averaging period). While under neutral conditions the wind forcing and wind direction are reasonably well defined, this is no longer the case as stratification becomes progressively more unstable ($-R_{if} > 1$). The prevalence of convective cells without a well-defined mean wind direction leads to an ill-defined coordinate system. Additionally, if significant veering/backing of wind with height is present, applying double rotation to each height individually will underestimate the contribution from directional shear to the Reynolds stresses. The influence of this choice on the results is further explored (Fig. \ref{fig:MetCrax_rotation}). 

First we explore, how well-defined the coordinate system is. As expected, the results for METCRAX II dataset highlight the increasing variability of wind direction within the $30min$ averaging window with increasing instability (Fig. \ref{fig:MetCrax_rotation}a), with standard deviation of wind direction reaching as much as $50^\circ$ at 3 m level. While this effect can influence the energy content of streamwise and spanwise variances in very unstable stratification, the observed increase of spanwise variance ratio (Fig. \ref{fig:Flat_var}) occurs at lower $R_{if}$ where the coordinate system is still well defined. 

Second, the veer of the mean wind could misalign the streamwise direction at different elevations. Thus, what would be considered spanwise variance at one level would have a streamwise component at another level, and transport can bring that streamwise variance to an elevation where it would be normal to the streamlines. To test this hypothesis we applied a uniform coordinate system across all heights, where the x direction at each height coincided with the mean wind direction at the lowest observational level. The estimated contribution of the spanwise shear production that is present in this configuration (Fig. \ref{fig:MetCrax_rotation}b), however, is multiple orders of magnitude smaller than other terms in the budgets and can thus be neglected. Additionally, no significant veering of the wind with height close to the ground, where the spanwise variance dominates, was observed in any of the datasets. 

Finally, another reason for the high spanwise variance could be the contribution of the Coriolis terms. However, the contribution of this term is negligible (Fig. \ref{fig:MetCrax_rotation}c) as can also be inferred by estimating its ratio relative to the dominant shear production component as $f_c/(d\overline{U}/dz) \sim f_c\kappa z/u_*$, which will be very small in almost all ABL regimes, with $u_*$ the friction velocity and $\kappa$ the von Kármán constant.

\subsection{Budget approach in the convective boundary layer}

Convective boundary layers (CBL) are characterized by the organization of turbulence into coherent structures, in which bottom-up and top-down processes exhibit marked differences \citep{MoengWyngaard1989}. CBL turbulence structure is therefore inherently non-local. 
This feature makes local closure models of single-point averaged conservation equations questionable \citep[e.g.,][]{Mishra2016}. This is particularly an issue when turbulent and pressure transports are parametrized. In fact, LES runs suggest a non-zero residual even when all budget terms are accounted for \citep[e.g.,][]{RotachHoltslag2025}. 

\begin{figure}
    \centering
        \includegraphics[width = 1\linewidth]{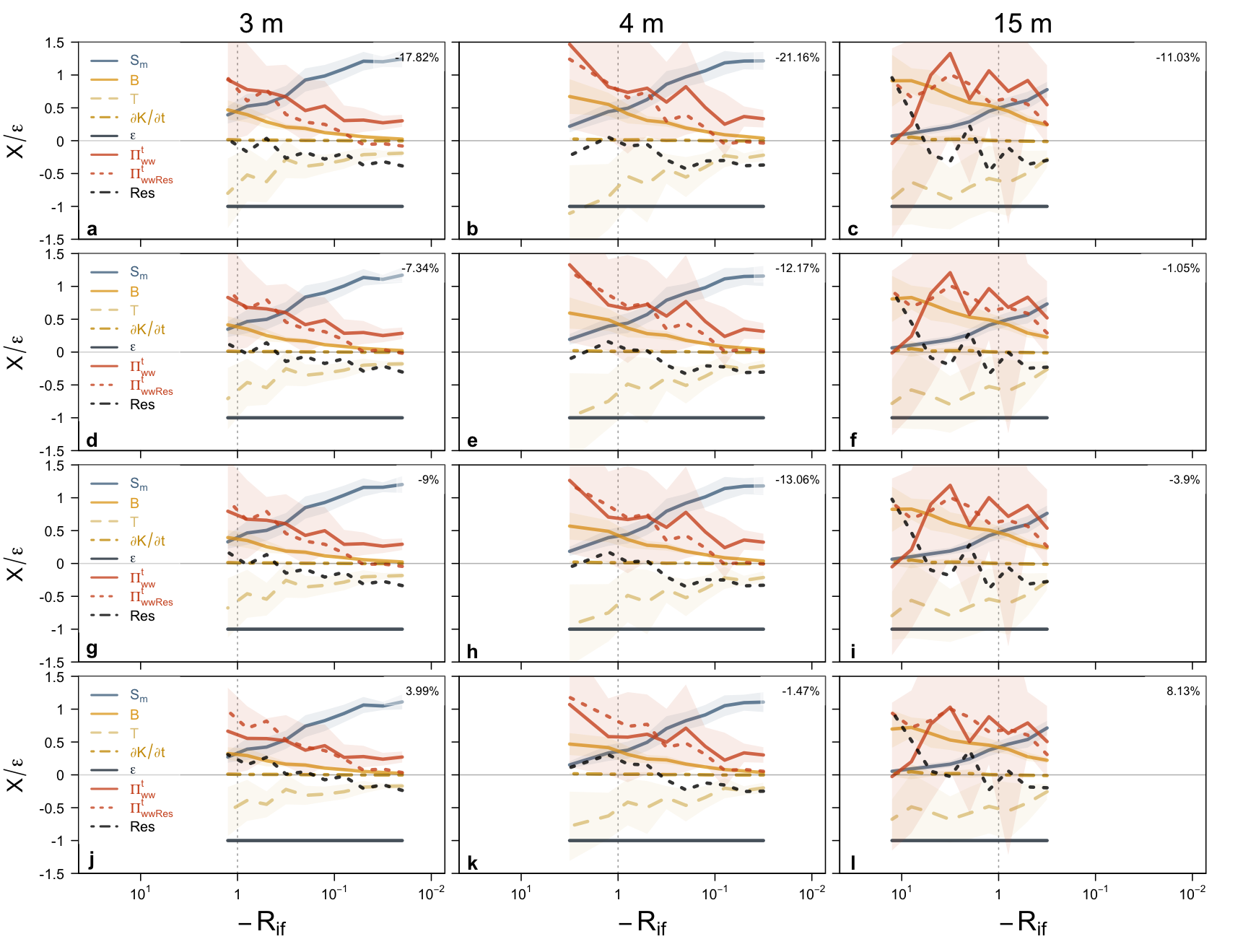}
    \caption{Normalized TKE budget terms for the (a,d,g,j) 3\,m, (b,e,h,k) 4\,m, and (c,f,i,l) 15\,m level of the M2HATS tower as a function of $R_{if}$, for TKE dissipation rate estimates as (a-c) $(1/2)(\varepsilon_u +\varepsilon_v)$, (d-f) $\varepsilon_u$, (g-i) $(1/2)(\varepsilon_{ucorr} +\varepsilon_{vcorr})$, and (j-l) $\varepsilon_{ucorr}$, where subscript $corr$ refers to the turbulence intensity correction. Budget terms and colours are the same as in Fig. \ref{fig:Flat_TKE}. Black dashed line corresponds to the total residual of TKE when pressure transport is directly measured. Numbers in the top right corner correspond to the mean residual as a percentage of the total dissipation.}
    \label{fig:M2HATS_TKE}
\end{figure}

Apart from pressure redistribution terms, all other processes in the turbulent stress budget are independently estimated from measurements for the M2HATS dataset. Still, estimating the budget terms from observations carries uncertainties already discussed (see Sect. \ref{sec:TKE}). One of the uncertainties stems from the estimation of the TKE dissipation rate itself. We therefore examine its influence on the pressure-transport term estimated as the residual of the TKE budget and compare it to the one directly measured (Fig. \ref{fig:M2HATS_TKE}). The results show that the pressure transport estimated as the residual of the TKE budget (dashed pink lines) and directly measured (full pink line) match at all heights. The exact match is a function of the dissipation rate chosen, though. If the $\varepsilon$ is computed from the streamwise dissipation rate accounting for turbulence intensity correction, the total residual of the TKE budget is less than 10\% at all examined heights, a value that is smaller than obtained from LES studies. To assess if the leftover 10\% residual stems from the horizontal terms or other non-local influences would require employing multi-point approaches based on networks of observational towers, Lagrangian approaches, or methods such as high-resolution LES that are outside of the scope of this study. 

\section{Conclusions}
\label{sec:conclusions}

The Reynolds stress conservation equations for a statically unstable atmospheric surface layer at very high Reynolds number were analysed using field measurements from multiple sites. All the sites chosen are flat and horizontally homogeneous, and experience a wide range of atmospheric instability conditions. In addition, the measurements interrogate many wall-normal distances that bound the atmospheric surface layer. Focus was placed on pressure-strain and the validity of common parametrizations used to model this term with the goal of understanding the drivers of Reynolds stress anisotropy.

The results show an expected decrease of the importance of streamwise velocity variance in the total TKE as the stratification becomes highly convective, but also highlight a consistent decrease of wall-normal velocity variance (below 20 m) and an increase of spanwise variance (up to 50 m)  
when the ASL shifts from neutral to intermediate instabilities $-R_{if} =[-0.1,1]$. The evaluation of the different budget terms shows that this behaviour can only be reproduced if, alongside production and dissipation mechanisms (equilibrium state), the turbulent and pressure transport are also taken into account. From the perspective of pressure-strain, the rapid isotropization of production and buoyancy terms are of added importance in the intermediate instability range, where the flow is shown to undergo a rapid distortion. 

The pressure transport is shown to be connected with the organization of turbulence into coherent structures, with a key role in driving the near-surface anisotropy, and explaining the dominant contribution of spanwise variance to the total turbulent kinetic energy, as well as the drop in wall-normal variance. In addition, the transition between the different types of coherent structures is shown to result in rapid distortion, particularly active in the transitional, dynamic-convective regime at intermediate instabilities, a process that has so far not been noted or studied. \cite{launder1975progress} and later others \citep{so1977model} highlight that rapid distortion has a strong influence over curved surfaces such as hilly terrain \citep[see][]{Mosso2025}, which lends itself to a conjecture that the streamline curvature of organized motions can play a similar role. Through this process, turbulence undergoes a rapid distortion that leads to increased anisotropy. Neglecting these processes (pressure transport and rapid distortion) could lead to an interpretation that is counter-intuitive, suggesting that a negative pressure-strain interaction in the dynamic-convective regime causes the proportion of energy in the wall-normal variance to drop as the instability increases. The recipient of the excess energy, however, is here shown to be the spanwise variance, at the expense of both the least energetic wall-normal variance, as well as the most energetic streamwise variance.
 
The importance of spanwise variance has so far not received due attention. Its observations or systematic analysis are generally lacking in many laboratory studies, perhaps due to unavoidable lateral confinement (in wind tunnels and flumes). The work here shows the spanwise variance originates from pressure-transport, which was historically deemed as minor \citep{KaimalFinnigan1994} on the one hand, and the damping, at the largest scale, of the streamwise variance by the shear and the wall-normal variance by wall blockage on the other. Thus spanwise variance is able to store energy at the largest scales. Recent work also showed that the behaviour of spanwise variance is a function of surface roughness \citep{Waterman2025}. Over flat terrain with small roughness elements, the spanwise variance has similar energy content to the streamwise variance \citep{Stiperski2023}. However, over vegetated canopies, the contribution of spanwise variance to the total TKE is consistently lower \citep[see,][]{Waterman2025}, thus potentially pointing to the role of roughness on coherent structures. Studies that directly observe turbulent pressure variations in all three directions, especially over complex surfaces, are needed to close some of these existing knowledge gaps. 

The results presented here offer guidance for future parametrization development. A key finding is that the pressure-strain parametrizations should not be applied to the total pressure term $\Pi_{ij}$ as proposed in few prior studies \citep{Zeman1981,Canuto2001}. The pressure transport term has a separate role in the Reynolds stress budgets, especially in stratified flows, and cannot be accounted for through the rapid pressure-strain parametrizations. Additionally, the results show that all of the rapid terms are important, although buoyancy has a more limited role (curves with $C^u_B$ = 0.3 and 0.6 show similar behaviour) \citep[see][]{Ding2018}. On the other hand, the isotropization of the production term representing vorticity plays a prominent role in pressure-strain interactions and should not be neglected. In fact, a simplified co-spectral budget did show a theoretical link between the numerical values of the von Kármán constant, the isotropization of production coefficient ($=3/5$ predicted from rapid distortion theory), the standard Rotta constant, and the Kolmogorov constant \citep{katul2013co}, pointing to the significance of reduced production in momentum fluxes.   Finally, even the more complex non-linear Rotta closure requires adjustments for wall blocking very close to the surface to correctly capture the observed energy partition.  Its significance in describing the mean velocity profile in the buffer layer of canonical smooth-wall bounded flows has been well established \citep{mccoll2016mean}, and the work here goes beyond to illustrate its significance in the mechanics of return to isotropy.   

\section*{Declaration of Interests}
The authors report no conflict of interest
 
\section*{Acknowledgements}
These results are part of a project that has received funding from the European Research Council (ERC) under the European Union’s Horizon 2020 research and innovation program (Grant agreement No. 101001691). 
GK acknowledges support from the U.S. National Science Foundation (NSF-AGS-2028633) and the U.S. Department of Energy (DE-SC0022072). EBZ acknowledges support from the National Oceanic and Atmospheric Administration (U.S. Department of Commerce grant no. NA18OAR4320123) and Princeton University through the Cooperative Institute for Modeling the Earth System. MC acknowledges support from the U.S. National Science Foundation (AGS-EAGER-2414424, CBET-2235750).
The computational results presented here have been produced (in part) using the LEO HPC infrastructure of the University of Innsbruck.


\appendix
\section{}\label{appendixClosureSchemes}
Different common closure schemes used to parametrize the pressure-strain terms are summarized in Table \ref{tab:tableClosureSchemes}, together with the respective parameters found in the literature. Some schemes parametrize the full pressure-velocity covariance, while others parametrize the pressure-strain terms only. This explains the differences in signs between the two sets of parametrizations.

\begin{table}
\small
\centering
\caption{Literature models of pressure-strain terms and associated constants}
\resizebox{\columnwidth}{!}{%
\begin{tabular}{lll}
\textbf{Author} & \textbf{Parametrization} & \textbf{Coefficients} \\
\midrule
\textbf{Zeman (1981)} & Pressure-velocity covariance \\
\midrule
Turbulence-turbulence interaction  & $\Pi_{ij}^T= 2\frac{C_1}{\tau} b_{ij}e$ 
& $C_1 = 2 - 6$\\
\arrayrulecolor{black!30}\midrule
Shear and vorticity term   & 
$\!\begin{aligned} 
\Pi_{ij}^S = & -\frac{4}{5}S_{ij}e \\ 
& -4\alpha_1(S_{ik}b_{kj} + S_{jk}b_{ki} -\frac{2}{3}S_{kl}b_{kl}\delta_{ij})e \\
& - 4\alpha_2(R_{ik}b_{kj} + R_{jk}b_{ki})e 
\end{aligned}$  &
$\!\begin{aligned}
& \alpha_1 = 3/7\\ & \alpha_2 = 0
\end{aligned}$ 
\\
\arrayrulecolor{black!30}\midrule
Buoyancy term  & $\Pi_{ij}^B= -\frac{3}{10}\frac{g}{\theta}(\overline{u_i'\theta'}\delta_{3j} + \overline{u_{k}'\theta'}\delta_{3i}- \frac{2}{3}\overline{u_{3}'\theta'}\delta_{ij})$ 
\\
\arrayrulecolor{black!100}\midrule
\textbf{Canuto et al. (2001)} & Pressure-velocity covariance \\
\arrayrulecolor{black!100}\midrule
Turbulence-turbulence interaction  & $\Pi_{ij}^T= \frac{2}{\tau_{p\nu}} b_{ij}$ & $\frac{\tau_{p\nu}}{\tau} = \frac{2}{5}\frac{\varepsilon}{2e}$ \\
\arrayrulecolor{black!30}\midrule
Shear and vorticity term   & 
$\!\begin{aligned} 
\Pi_{ij}^S = & -\frac{4}{5}S_{ij}e \\ 
& -\alpha_1(S_{ik}b_{kj} + S_{jk}b_{ik} -\frac{2}{3}S_{kl}b_{kl}\delta_{ij}) \\
& -\alpha_2(R_{ik}b_{kj} + R_{jk}b_{ik}) 
\end{aligned}$  &

$\!\begin{aligned}
& \alpha_1 = 6\frac{1}{10}(1+\frac{4}{5}\sqrt{0.64}) \\ & \alpha_2 =\frac{2}{3}[2-\frac{7}{10}(1+\frac{4}{5}\sqrt{0.64})]
\end{aligned}$ 
\\
\arrayrulecolor{black!30}\midrule
Buoyancy term  & $\Pi_{ij}^B= + (1 - \beta_{5}(\lambda_{i}h_{j}+\lambda_{j}h_{i} -\frac{2}{3}\lambda{j}h_{k}\delta_{ij}]$ 
& $\beta_5=0.48$\\
\arrayrulecolor{black!100}\midrule
\textbf{Heinze et al. (2016)} & Pressure-strain term \\
\arrayrulecolor{black!100}\midrule
Turbulence-turbulence interaction  & $\Pi_{ij}^T= -\frac{C_T^u}{\tau} b_{ij}e$ 
& $C_T^u = 1.5 - 1.8$ (HL, 2011); $1 - 3$ (Zeman 1981)\\
\arrayrulecolor{black!30}\midrule
Shear and vorticity term   & 
$\!\begin{aligned} 
\Pi_{ij}^S = & \frac{4}{5}S_{ij}e \\ 
& + C_{S1}^u(b_{jk}S_{ik} + b_{ik}S_{jk} -\frac{2}{3}b_{kl}S_{kl}\delta_{ij})e \\
& + C_{S2}^u(b_{ik}R_{jk} + b_{jk}R_{ik})e 
\end{aligned}$  &
$\!\begin{aligned}
& C_{S1}^u = \frac{12}{7};\frac{3}{5}\\ & C_{S2}^u =0; \frac{3}{5}
\end{aligned}$ 
\\
\arrayrulecolor{black!30}\midrule
Buoyancy term  & $\Pi_{ij}^B= -C_{B}^u\frac{g}{\overline{\theta}}(\overline{u_j'\theta'}\delta_{i3} + \overline{u_{i}'\theta'}\delta_{j3}- \frac{2}{3}\overline{u_{k}'\theta'}\delta_{k3})$ 
& $C_{B}^u=0.3 - 0.6; \frac{3}{10}$ for isotropic turbulence \\

\arrayrulecolor{black!100}\bottomrule
\end{tabular}
}
\label{tab:tableClosureSchemes}
\vspace{-4mm}
\end{table}


\section{}\label{appendixFullBudgetEquations}
The full set of budget equations for half-variances used to derive the extended model in Eqs \ref{eq:FullX} - \ref{eq:FullZ}:

\begin{align}
  \overline{u'^2}:\quad & \underbrace{-\overline{u'w'}\frac{\partial\overline{U}}{\partial
    z}}_{S_m}
  \underbrace{-\frac{1}{2}\frac{\partial\overline{w'u'^2}}{\partial z}}_{T_u}  -\mathrm{\varepsilon_{u}} 
  \nonumber \\
  & \underbrace{-\frac{c \varepsilon}{e} \left[\overline{u'^2} -\frac{2}{3}e a_u\right] + C_{B}^u\frac{1}{3}\frac{g}{\overline{\theta}}\overline{w'\theta'} + \overline{u'w'}\frac{\partial\overline{U}}{\partial
    z} \frac{1}{2}\left[C_{S1}^u\frac{1}{3} + C_{S2}^u \right]}_{\Pi_{uu}}=0,
  \label{eq:ReynoldsX}
\end{align}

\begin{align}
 \overline{v'^2}: \quad & 
  \underbrace{-\frac{1}{2}\frac{\partial\overline{w'v'^2}}{\partial z}}_{T_v}  -\mathrm{\varepsilon_{v}} 
  \nonumber \\
  & \underbrace{-\frac{c \varepsilon}{e} \left[\overline{v'^2} -\frac{2}{3}e a_v\right] + C_{B}^u\frac{1}{3}\frac{g}{\overline{\theta}}\overline{w'\theta'} + \overline{u'w'}\frac{\partial\overline{U}}{\partial
    z} \frac{1}{2}\left[-C_{S1}^u\frac{2}{3} \right]}_{\Pi_{vv}}=0,
  \label{eq:ReynoldsY}
\end{align}

\begin{align}
 \overline{w'^2}: \quad & 
  \underbrace{\frac{g}{\overline{\theta}}\overline{w'\theta_v'}}_{B}
  \underbrace{-\frac{1}{2}\frac{\partial\overline{w'^3}}{\partial z}}_{T_w}  -\mathrm{\varepsilon_{w}} 
  \nonumber \\
  & \underbrace{-\frac{c \varepsilon}{e} \left[\overline{w'^2} -\frac{2}{3}e a_w\right] - C_{B}^u\frac{2}{3}\frac{g}{\overline{\theta}}\overline{w'\theta'} + \overline{u'w'}\frac{\partial\overline{U}}{\partial
    z} \frac{1}{2}\left[C_{S1}^u\frac{1}{3} - C_{S2}^u \right]}_{\Pi_{ww}}=0,
  \label{eq:ReynoldsZ}
\end{align}

Here, the dissipation rates of each half-variance are defined as $\varepsilon_u,\varepsilon_v, \varepsilon_w$, which under the assumption of isotropic dissipation will be $\varepsilon/3$.  

\section{}\label{appendixAnisotropicDissipation}

The extended model that, apart from transport and rapid pressure-strain terms, also includes the anisotropy of dissipation term. The persistence of such fine-scaled dissipation anisotropy at high Reynolds number could be modelled as 
\begin{equation}
\varepsilon_u + \varepsilon_v = \alpha \varepsilon_w,
\end{equation}
where $\alpha = 2$ recovers the isotropic dissipation.

\begin{align}
  \frac{\overline{u'^2}}{e}= \nonumber \\& \frac{1}{c \varepsilon} \left[ \left( \frac{\alpha}{\alpha + 1} + \frac{1}{\alpha + 1} R_{if} \right)S_m + \frac{1}{3}C_{B}^uB - \left(\frac{1}{3} C^u_{S1} + C^u_{S2} \right)\frac{S_m}{2} +  \frac{\left( T_w + T_v -2T_u \right)}{3} \right]+ \frac{2}{3}a_u, \label{eq:FullX_anisdis} \\
  \frac{\overline{v'^2}}{e}= \nonumber \\ & \frac{1}{c \varepsilon} \left[ \left( -\frac{1}{\alpha + 1} + \frac{1}{\alpha + 1} R_{if} \right)S_m + \frac{1}{3}C_{B}^uB + \frac{2}{3}C^u_{S1}\frac{S_m}{2} + \frac{\left(T_w +T_u -2T_v\right)}{3} \right] + \frac{2}{3}a_v, 
  \label{eq:FullY_anisdis} \\
  \frac{\overline{w'^2}}{e}= \nonumber \\ & \frac{1}{c \varepsilon} \left[ \left( -\frac{1}{\alpha + 1} - \frac{\alpha}{\alpha + 1} R_{if} \right)S_m - \frac{2}{3}C_{B}^u B - \left(\frac{1}{3} C^u_{S1} - C^u_{S2} \right)\frac{S_m}{2} + \frac{\left( T_u + T_v -2T_w \right)}{3} \right] + \frac{2}{3}a_w. \label{eq:FullZ_anisdis} 
\end{align}
To estimate this parameter from data, one could use the dissipation rates obtained from the spectra of three velocity components, where $\varepsilon_w$ is oftentimes shown to be smaller than those of the horizontal components.  


\section{}\label{appendixOtherFigures}

Figures from other datasets supporting the results presented for one dataset in the main manuscript text. 

\begin{figure}
    \centering
        \includegraphics[width = 0.8\linewidth]{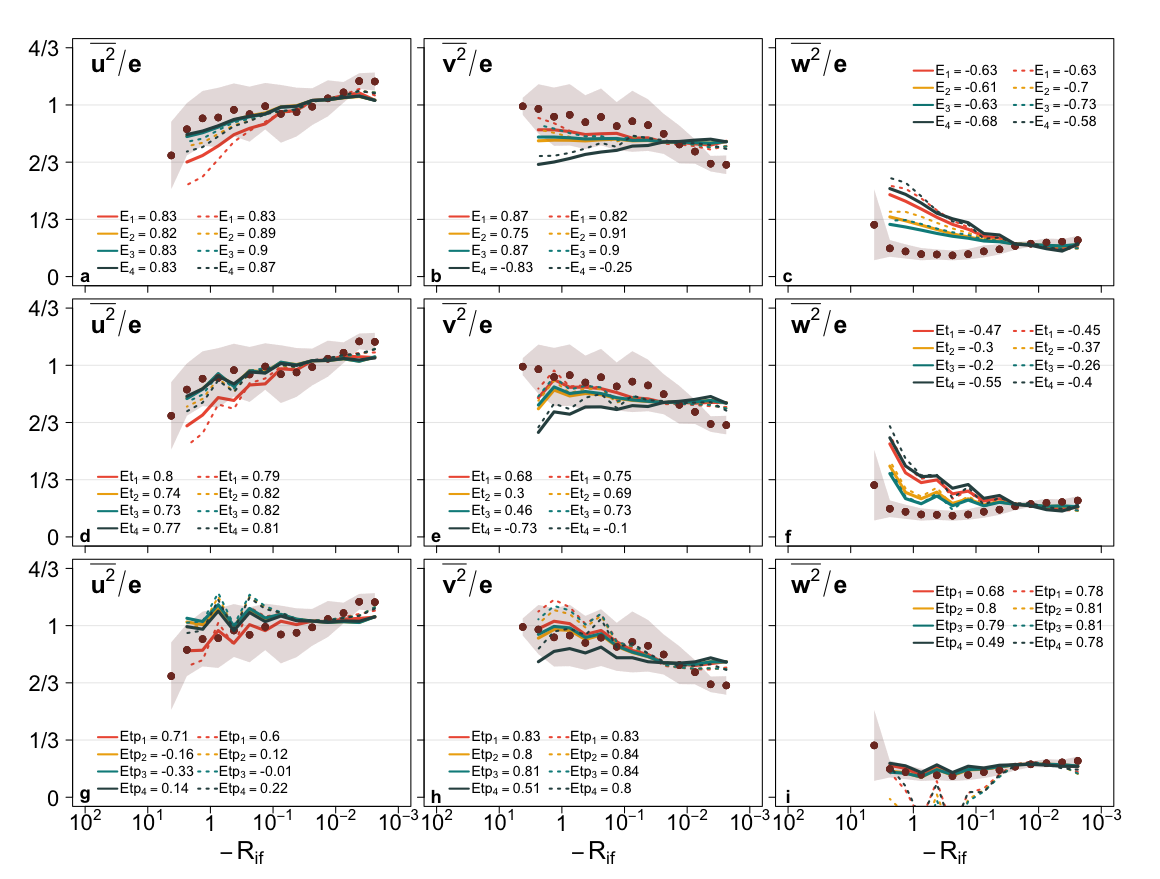}
    \caption{Same as Fig. \ref{fig:Cabauw_Model} but for the 3 m level of the METCRAX II site.}
    \label{fig:Metcrax_Model}
\end{figure}

\begin{figure}
    \centering
        \includegraphics[width = 0.8\linewidth]{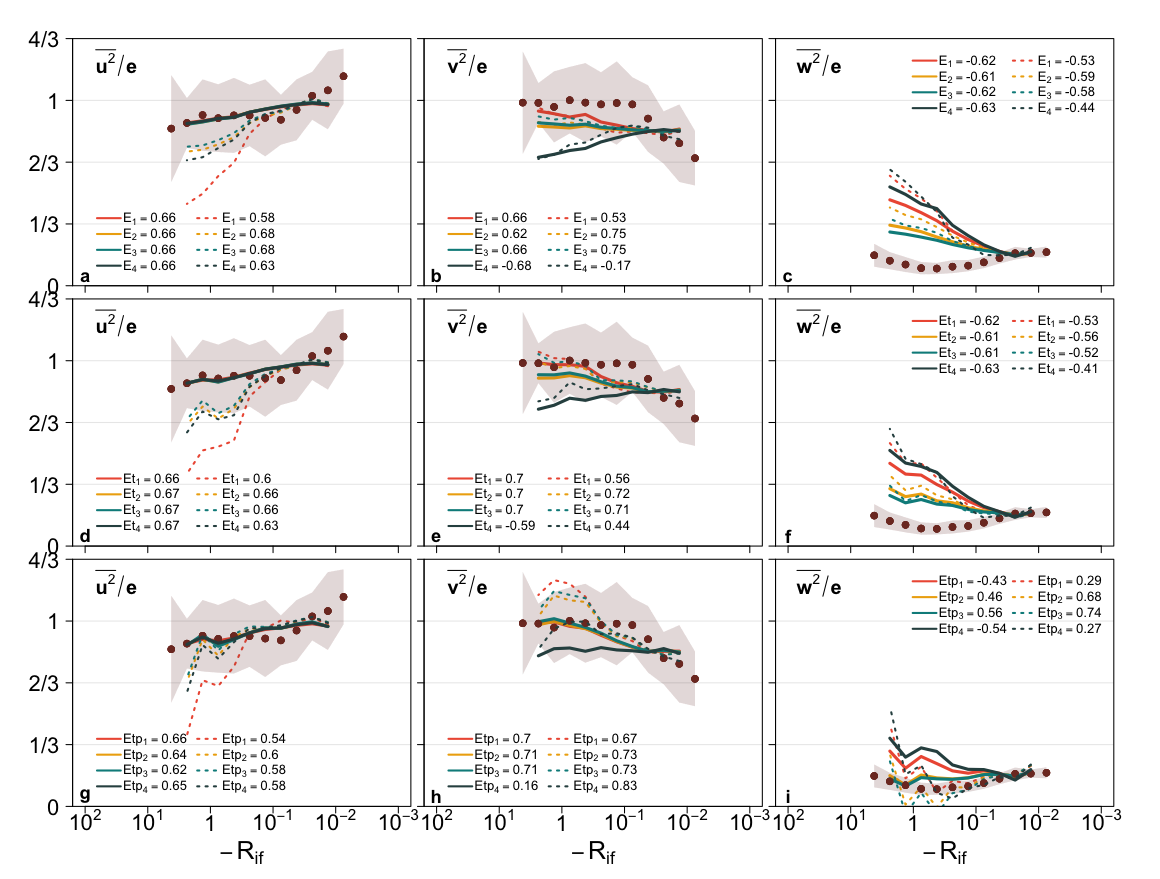}
    \caption{Same as Fig. \ref{fig:Cabauw_Model} but for the 3 m level of the M2HATS tower. Unlike in Figs. \ref{fig:Cabauw_Model} and \ref{fig:Metcrax_Model}, however, the directly measured pressure transport term is used in panels (g - i) instead of the one estimated as the residual of the TKE budget.}
    \label{fig:M2HATS_Model}
\end{figure}

Figures \ref{fig:Metcrax_Model} and \ref{fig:M2HATS_Model} show the performance of the different models (see table \ref{tab:tableModels}) in describing the velocity variance ratios of the lowest observational level of the METCRAX II and the M2HATS sites. The results are consistent with those presented for Cabauw (Fig. \ref{fig:Cabauw_Model}).  A notable exception is that the correct behaviour of both the wall normal and spanwise variance is only reproduced when the pressure transport is included in the budget (\textbf{Model Etp}).

\begin{figure}
    \centering
        \includegraphics[width = 0.8\linewidth]{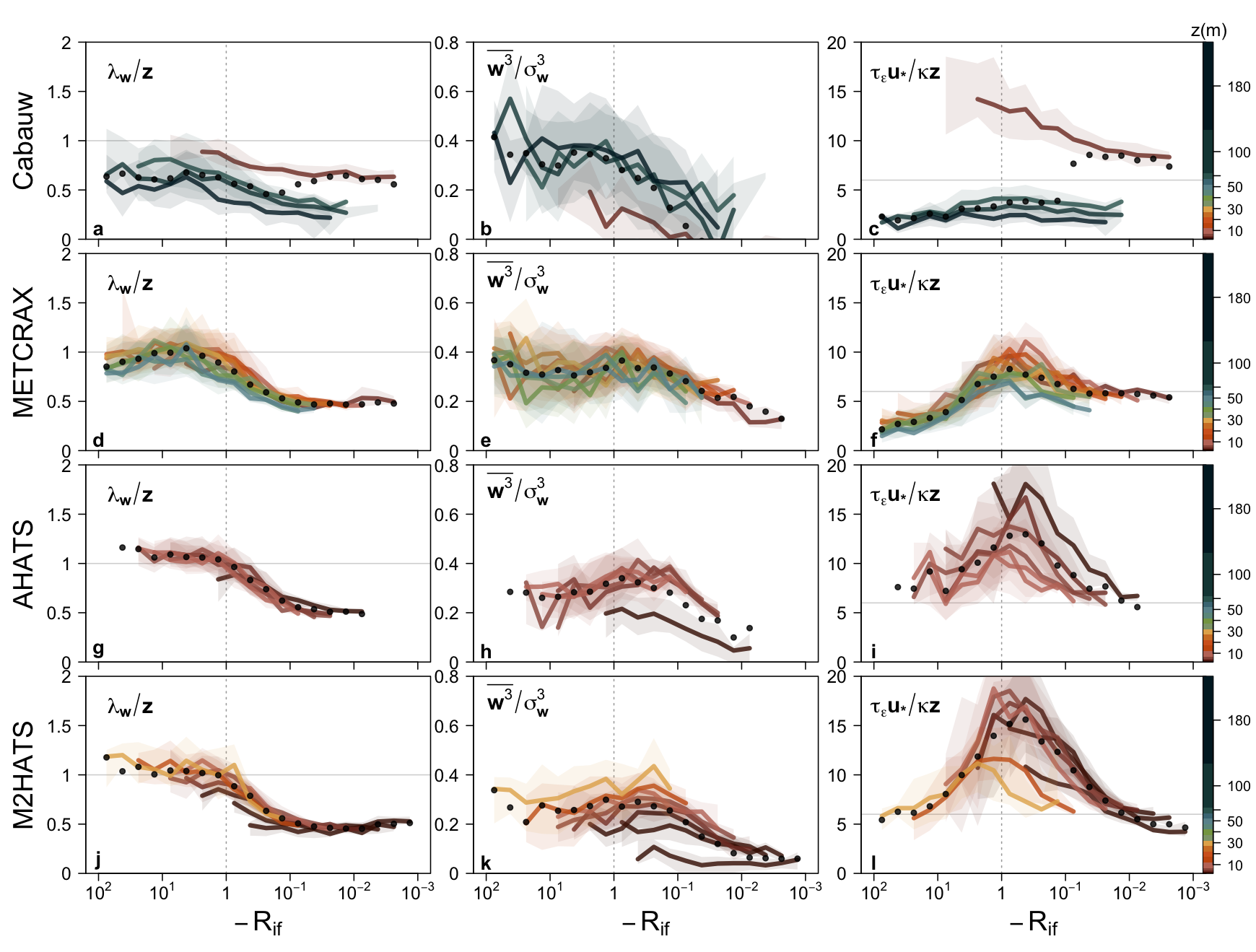}
    \caption{Same as Fig. \ref{fig:MetCrax_RDT} but for all datasets.}
    \label{fig:Flat_RDT}
\end{figure}

Figure \ref{fig:Flat_RDT} highlights the same behaviour of non-dimensional Eulerian integral length scale, skewness of wall-normal velocity, and normalized rapid distortion timescale at low observational heights (brown lines) of all the datasets as already found for METCRAX. As expected, the difference is the upper levels of the Cabauw dataset (60 - 180m) for which the variance ratios are reasonably predicted by the reduced model with added wall blocking. 

\begin{figure}
    \centering
        \includegraphics[width = 0.8\linewidth]{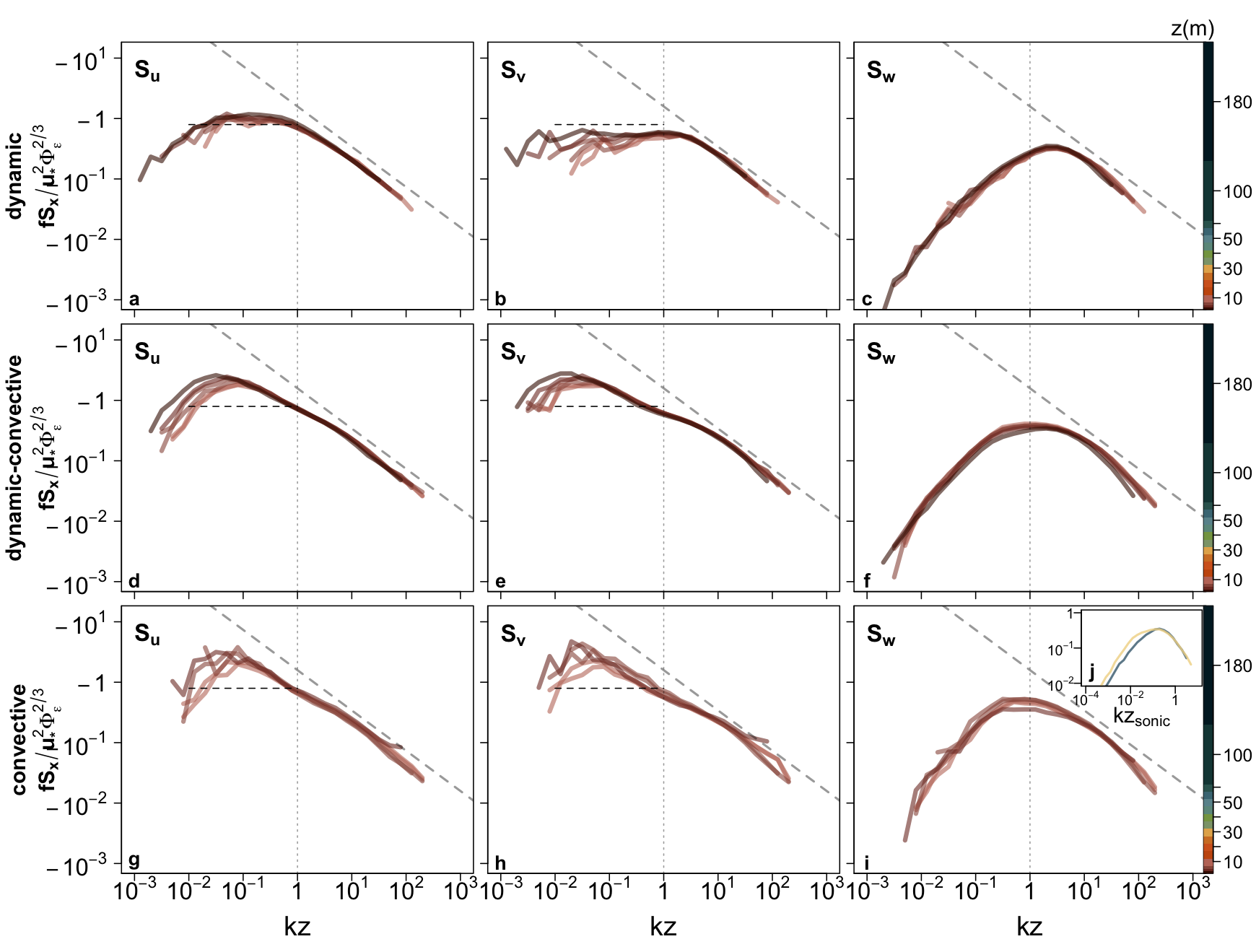}
    \caption{Same as Fig. \ref{fig:MetCrax_Spectra} but for the AHATS dataset.}
    \label{fig:AHATS_Spectra}
\end{figure}

\begin{figure}
    \centering
        \includegraphics[width = 0.8\linewidth]{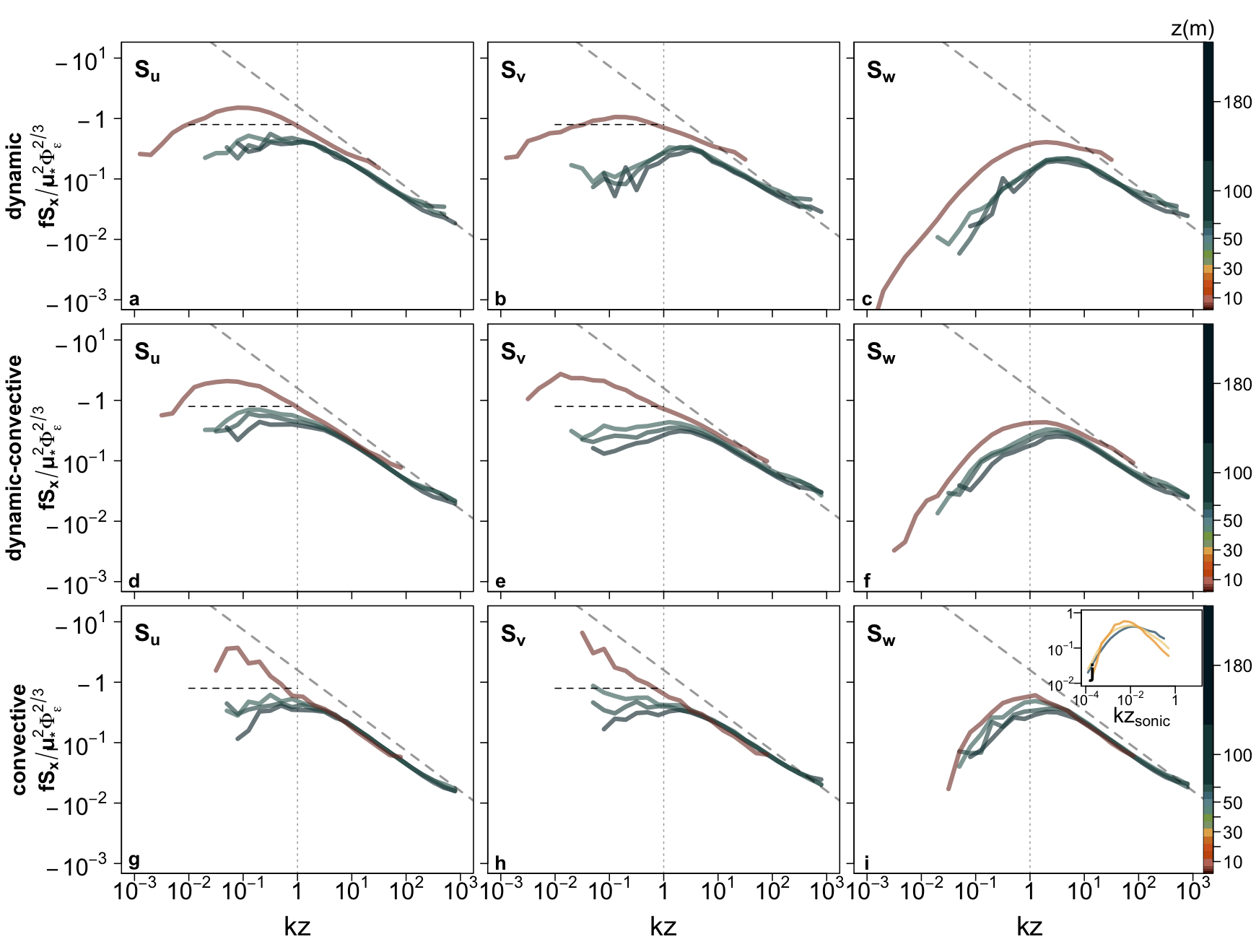}
    \caption{Same as Fig. \ref{fig:MetCrax_Spectra} but for the Cabauw tower.}
    \label{fig:Cabauw_Spectra}
\end{figure}

The scaled spectra in the three regimes (Figs. \ref{fig:AHATS_Spectra}, 
\ref{fig:Cabauw_Spectra}) for AHATS and Cabauw show consistent low frequency and high frequency spectral ranges to those in the main text. 

\bibliographystyle{jfm}
\bibliography{bibliography}

@BOOK{Aubinet2012,
    author = "Marc Aubinet and Timo Vesala and Dario Papale",
    title = "Eddy Covariance: A Practical Guide to Measurement and Data Analysis",
    publisher = "Springer",
    year = "2012",
}

@Article{Biltoft2001,
author={Biltoft, C. A.},
title={Some Thoughts On Local Isotropy And The 4/3 Lateral To Longitudinal Velocity Spectrum Ratio},
journal={Boundary-Layer Meteorology},
year={2001},
month={Sep},
day={01},
volume={100},
number={3},
pages={393-404},
issn={1573-1472},
doi={10.1023/A:1019289915930},
url={https://doi.org/10.1023/A:1019289915930}
}

@article{banerjee2013logarithmic,
  title={Logarithmic scaling in the longitudinal velocity variance explained by a spectral budget},
  author={Banerjee, T. and Katul, G.G.},
  journal={Physics of Fluids},
  volume={25},
  number={12},
  year={2013},
  publisher={AIP Publishing}
}

@article{Banerjee2007, 
year = {2007}, 
title = {Presentation of anisotropy properties of turbulence, invariants versus eigenvalue approaches}, 
author = {Banerjee, S. and Krahl, R. and Durst, F. and Zenger, Ch.}, 
journal = {Journal of Turbulence}, 
issn = {1468-5248}, 
doi = {10.1080/14685240701506896}, 
pages = {N32}, 
volume = {8}
}

@article{BeljaarsBosveld1997,
author = {Beljaars, Anton C. M. and Bosveld, Fred C.},
title = {Cabauw Data for the Validation of Land Surface Parameterization Schemes},
journal = {Journal of Climate},
volume = {10},
number = {6},
pages = {1172--1193},
year = {1997},
doi = {10.1175/1520-0442(1997)010<1172:CDFTVO>2.0.CO;2}
}

@article{BouZeid2018,
	Author = {Bou-Zeid, Elie and Gao, Xiang and Ansorge, Cedrick and Katul, Gabriel G},
	Journal = {Journal of Fluid Mechanics},
	Pages = {61--78},
	Title = {{On the role of return to isotropy in wall-bounded turbulent flows with buoyancy}},
	Volume = {856},
	Year = {2018},
	doi = {10.1017/jfm.2018.693}}

@Article{Cuxart2002,
author={Cuxart, J.
and Morales, G.
and Terradellas, E.
and Yag{\"u}e, C.},
title={Study of Coherent Structures and Estimation of the Pressure Transport Terms for the Nocturnal Stable Boundary Layer},
journal={Boundary-Layer Meteorology},
year={2002},
month={Nov},
day={01},
volume={105},
number={2},
pages={305-328},
issn={1573-1472},
doi={10.1023/A:1019974021434}
}

@article{Chamecki2004,
    author = {Chamecki, M. and Dias, N. L.},
    title = {The local isotropy hypothesis and the turbulent kinetic energy dissipation rate in the atmospheric surface layer},
    journal = {Quarterly Journal of the Royal Meteorological Society},
    volume = {130},
    number = {603},
    pages = {2733--2752},
    year = {2004}}

@article {Canuto2001,
      author = "V. M. Canuto and A. Howard and Y. Cheng and M. S. Dubovikov",
      title = "Ocean Turbulence. {P}art {I}: One-Point Closure Model—Momentum and Heat Vertical Diffusivities",
      journal = "Journal of Physical Oceanography",
      year = "2001",
      publisher = "American Meteorological Society",
      address = "Boston MA, USA",
      volume = "31",
      number = "6",
      doi = "10.1175/1520-0485(2001)031<1413:OTPIOP>2.0.CO;2",
      pages=      "1413 - 1426",
}

@article{Chowdhuri2024,
  title = {Quantifying small-scale anisotropy in turbulent flows},
  author = {Chowdhuri, S. and Banerjee, T.},
  journal = {Physical Review Fluids},
  volume = {9},
  issue = {7},
  pages = {074604},
  numpages = {18},
  year = {2024},
  month = {Jul},
  publisher = {American Physical Society},
  doi = {10.1103/PhysRevFluids.9.074604},
  url = {https://link.aps.org/doi/10.1103/PhysRevFluids.9.074604}
}

@article{Charrondiere2024,
author = {Charrondière, Claudine and Stiperski, Ivana},
title = {Spectral scaling of unstably stratified atmospheric flows: Turbulence anisotropy and the low-frequency spread},
journal = {Quarterly Journal of the Royal Meteorological Society},
year = {2024},
volume = {150},
number = {764},
pages = {4196-4216},
doi = {https://doi.org/10.1002/qj.4811}
}

@article{CHOI_LUMLEY_2001, title={The return to isotropy of homogeneous turbulence}, volume={436}, 
DOI={10.1017/S002211200100386X}, journal={Journal of Fluid Mechanics}, 
author={Choi, K.S and Lumley, J.L.}, 
year={2001}, 
pages={59–84}}

@article{Ding2018,
  title={Investigation of the pressure--strain-rate correlation and pressure fluctuations in convective and near neutral atmospheric surface layers},
  author={Ding, Mengjie and Nguyen, Khuong X and Liu, Shuaishuai and Otte, Martin J and Tong, Chenning},
  journal={Journal of Fluid Mechanics},
  volume={854},
  pages={88--120},
  year={2018},
  publisher={Cambridge University Press}
}

@Article{Freire2019,
author={Freire, Livia S.
and Dias, Nelson L.
and Chamecki, Marcelo},
title={Effects of Path Averaging in a Sonic Anemometer on the Estimation of Turbulence-Kinetic-Energy Dissipation Rates},
journal={Boundary-Layer Meteorology},
year={2019},
month={Oct},
day={01},
volume={173},
number={1},
pages={99-113},
issn={1573-1472},
doi={10.1007/s10546-019-00453-4},
url={https://doi.org/10.1007/s10546-019-00453-4}
}

@article{Finnigan2020,
  title={Boundary-layer flow over complex topography},
  author={Finnigan, John and Ayotte, Keith and Harman, Ian and Katul, Gabriel and Oldroyd, Holly and Patton, Edward and Poggi, Davide and Ross, Andrew and Taylor, Peter},
  journal={Boundary-Layer Meteorology},
  volume={177},
  number={2},
  pages={247--313},
  year={2020},
  publisher={Springer}
}

@article{FokenWichura1996,
title = "Tools for quality assessment of surface-based flux measurements",
journal = "Agricultural and Forest Meteorology",
volume = "78",
number = "1",
pages = "83--105",
year = "1996",
note = "",
issn = "0168-1923",
author = "Th. Foken and B. Wichura",
doi = "10.1016/0168-1923(95)02248-1",
}

@article{GibsonLaunder1978, title={Ground effects on pressure fluctuations in the atmospheric boundary layer}, volume={86}, DOI={10.1017/S0022112078001251}, number={3}, journal={Journal of Fluid Mechanics}, author={Gibson, M. M. and Launder, B. E.}, year={1978}, pages={491–511}}

@Article{Goger2018,
    author="Goger, Brigitta
    and Rotach, Mathias W.
    and Gohm, Alexander
    and Fuhrer, Oliver
    and Stiperski, Ivana
    and Holtslag, Albert A. M.",
    title="The Impact of Three-Dimensional Effects on the Simulation of Turbulence Kinetic Energy in a Major Alpine Valley",
    journal="Boundary-Layer Meteorology",
    year="2018",
    month="Jul",
    day="01",
    volume="168",
    number="1",
    pages="1--27"}

@article{Hogstrom96,
  title={Review of some basic characteristics of the atmospheric surface layer},
  author={H{\"o}gstr{\"o}m, ULF},
  journal={Boundary-Layer Meteorology},
  volume={78},
  number={3},
  pages={215--246},
  year={1996},
  publisher={Springer},
doi = {10.1007/BF00120937}
}

@article{Homan2024,
  title = {Reynolds stress decay modeling informed by anisotropically forced homogeneous turbulence},
  author = {Homan, Ty and Shende, Omkar B. and Mani, Ali},
  journal = {Phys. Rev. Fluids},
  volume = {9},
  issue = {9},
  pages = {094608},
  numpages = {24},
  year = {2024},
  publisher = {American Physical Society},
  doi = {10.1103/PhysRevFluids.9.094608}
}

@Article{Hutchins2012,
author={Hutchins, Nicholas
and Chauhan, Kapil
and Marusic, Ivan
and Monty, Jason
and Klewicki, Joseph},
title={Towards Reconciling the Large-Scale Structure of Turbulent Boundary Layers in the Atmosphere and Laboratory},
journal={Boundary-Layer Meteorology},
year={2012},
month={Nov},
day={01},
volume={145},
number={2},
pages={273-306},
issn={1573-1472},
doi={10.1007/s10546-012-9735-4},
url={https://doi.org/10.1007/s10546-012-9735-4}
}

@article{Heinze2016,
author = {Heinze, Rieke and Mironov, Dmitrii and Raasch, Siegfried},
title = {Analysis of pressure-strain and pressure gradient-scalar covariances in cloud-topped boundary layers: A large-eddy simulation study},
journal = {Journal of Advances in Modeling Earth Systems},
volume = {8},
number = {1},
pages = {3-30},
doi = {https://doi.org/10.1002/2015MS000508},
year = {2016}
}

@article{hanjalic1972reynolds,
  title={A {R}eynolds stress model of turbulence and its application to thin shear flows},
  author={Hanjali{\'c}, Kemal and Launder, Brian E},
  journal={Journal of Fluid Mechanics},
  volume={52},
  number={4},
  pages={609--638},
  year={1972},
  publisher={Cambridge University Press}
}

@article{hanjalic2021reassessment,
  title={Reassessment of modeling turbulence via {R}eynolds averaging: A review of second-moment transport strategy},
  author={Hanjali{\'c}, K and Launder, BE},
  journal={Physics of Fluids},
  volume={33},
  number={9},
  year={2021},
  publisher={AIP Publishing}
}

@article{huang2022profiles,
  title={Profiles of high-order moments of longitudinal velocity explained by the random sweeping decorrelation hypothesis},
  author={Huang, K.Y. and Katul, G.G.},
  journal={Physical Review Fluids},
  volume={7},
  number={4},
  pages={044603},
  year={2022},
  publisher={APS}
}

@article{HuntGraham1978, 
author={Hunt, J. C. R. and Graham, J. M. R.},
title={Free-stream turbulence near plane boundaries}, volume={84}, 
DOI={10.1017/S0022112078000130}, 
number={2}, 
journal={Journal of Fluid Mechanics}, 
publisher={Cambridge University Press},  
year={1978}, 
pages={209–235}}

@article{hsieh1997dissipation,
  title={Dissipation methods, {T}aylor's hypothesis, and stability correction functions in the atmospheric surface layer},
  author={Hsieh, C-I and Katul, G.G.},
  journal={Journal of Geophysical Research: Atmospheres},
  volume={102},
  number={D14},
  pages={16391--16405},
  year={1997},
  publisher={Wiley Online Library}
}

@article{Jacobitz1999,
  title = {On the Shear Number Effect in Stratified Shear Flow},
  volume = {13},
  ISSN = {1432-2250},
  url = {http://dx.doi.org/10.1007/s001620050114},
  DOI = {10.1007/s001620050114},
  number = {3},
  journal = {Theoretical and Computational Fluid Dynamics},
  publisher = {Springer Science and Business Media LLC},
  author = {Jacobitz,  F.G. and Sarkar,  S.},
  year = {1999},
  month = aug,
  pages = {171–188}
}

@article{KassinosReynolds2001, 
title={One-point turbulence structure tensors}, 
author={Kassinos, S. C. and Reynolds, W. C. and Rogers, M. M.}, 
volume={428}, 
DOI={10.1017/S0022112000002615}, 
journal={Journal of Fluid Mechanics}, year={2001}, 
pages={213–248}}

@article{katul1996inactive,
  title={The inactive eddy motion and the large-scale turbulent pressure fluctuations in the dynamic sublayer},
  author={Katul, G.G. and Albertson, J.D. and Hsieh, C-I and Conklin, P.S and Sigmon, J.T and Parlange, M.B. and Knoerr, K.R.},
  journal={Journal of the Atmospheric Sciences},
  volume={53},
  number={17},
  pages={2512--2524},
  year={1996},
  publisher={[Boston, etc.] American Meteorological Society.}
}

@article{katul1997energy,
  title={Energy-inertial scale interactions for velocity and temperature in the unstable atmospheric surface layer},
  author={Katul, G. and Hsieh, C-I and Sigmon, J.},
  journal={Boundary-Layer Meteorology},
  volume={82},
  pages={49--80},
  year={1997},
  publisher={Springer}
}

@article{katul1998theoretical,
  title={A theoretical and experimental investigation of energy-containing scales in the dynamic sublayer of boundary-layer flows},
  author={Katul, G. and Chu, C-R},
  journal={Boundary-Layer Meteorology},
  volume={86},
  number={2},
  pages={279--312},
  year={1998},
  publisher={Springer}
}

@article{katul2012existence,
  title={Existence of k$^{-1}$ power-law scaling in the equilibrium regions of wall-bounded turbulence explained by {H}eisenberg's eddy viscosity},
  author={Katul, G.G. and Porporato, A. and Nikora, V.},
  journal={Physical Review E—Statistical, Nonlinear, and Soft Matter Physics},
  volume={86},
  number={6},
  pages={066311},
  year={2012},
  publisher={APS}
}

@article{katul2013co,
  title={Co-spectrum and mean velocity in turbulent boundary layers},
  author={Katul, Gabriel G and Porporato, Amilcare and Manes, Costantino and Meneveau, Charles},
  journal={Physics of Fluids},
  volume={25},
  number={9},
  pages={091702},
  year={2013},
  publisher={AIP Publishing}
}

@article{Kolmogorov1941a,
    author = {Kolmogorov, A.N.},
    title = {Dissipation of energy in locally isotropic turbulence},
    journal = {Dokl. Akad. Nauk SSSR},
    volume = {32},
    pages = {19-21},
    year = {1941}}

@article{KaderYaglom1990, 
title={Mean fields and fluctuation moments in unstably stratified turbulent boundary layers}, 
volume={212}, DOI={10.1017/S0022112090002129}, 
journal={Journal of Fluid Mechanics}, 
publisher={Cambridge University Press}, 
author={Kader, B. A. and Yaglom, A. M.}, year={1990}, pages={637–662}}

@incollection{kader1991spectra,
  title={Spectra and correlation functions of surface layer atmospheric turbulence in unstable thermal stratification},
  author={Kader, BA and Yaglom, AM},
  booktitle={Turbulence and Coherent Structures: Selected Papers from “Turbulence 89: Organized Structures and Turbulence in Fluid Mechanics”, Grenoble, 18--21 September 1989},
  pages={387--412},
  year={1991},
  publisher={Springer}
}

@book{KaimalFinnigan1994,
   title = {Atmospheric boundary layer flows: their structure and measurement},
   author = {JC Kaimal and JJ Finnigan},
   year = {1994},
   publisher = {Oxford University Press},
   address = {Oxford, UK}}

@book{LumleyPanofsky1964,
    author = {Lumley, J L and Panofsky, H A},
    title = {The structure of atmospheric turbulence},
    publisher = {Wiley, New York},
    year = {1964}
}

@article{LumleyNewman1977,
  title={The return to isotropy of homogeneous turbulence},
  author={Lumley, John L and Newman, Gary R},
  journal={Journal of Fluid Mechanics},
  volume={82},
  number={1},
  pages={161--178},
  year={1977},
  publisher={Cambridge University Press},
doi = {10.1017/S0022112077000585}
}

@article{Li2011,
  title = {Coherent Structures and the Dissimilarity of Turbulent Transport of Momentum and Scalars in the Unstable Atmospheric Surface Layer},
  volume = {140},
  ISSN = {1573-1472},
  url = {http://dx.doi.org/10.1007/s10546-011-9613-5},
  DOI = {10.1007/s10546-011-9613-5},
  number = {2},
  journal = {Boundary-Layer Meteorology},
  publisher = {Springer Science and Business Media LLC},
  author = {Li,  Dan and Bou-Zeid,  Elie},
  year = {2011},
  month = apr,
  pages = {243–262}
}

@article{Lin2000,
    author = {Lin, Ching-Long},
    title = {Local pressure-transport structure in a convective atmospheric boundary layer},
    journal = {Physics of Fluids},
    volume = {12},
    number = {5},
    pages = {1112-1128},
    year = {2000},
    month = {05},
    issn = {1070-6631},
    doi = {10.1063/1.870365}
}

@article{LiHatchinsMarusic2022, 
title={Scale-dependent inclination angle of turbulent structures in stratified atmospheric surface layers}, 
author={Li, Xuebo and Hutchins, Nicholas and Zheng, Xiaojing and Marusic, Ivan and Baars, Woutijn J.},
volume={942}, 
DOI={10.1017/jfm.2022.403}, 
journal={Journal of Fluid Mechanics},  
year={2022}, 
pages={A38}}

@article{launder1975progress,
  title={Progress in the development of a {R}eynolds-stress turbulence closure},
  author={Launder, B.E. and Reece, G. and Rodi, W.},
  journal={Journal of Fluid Mechanics},
  volume={68},
  number={3},
  pages={537--566},
  year={1975},
  publisher={Cambridge University Press}
}

@article{lenschow1994long,
  title={How long is long enough when measuring fluxes and other turbulence statistics?},
  author={Lenschow, DH and Mann, Jakob and Kristensen, Leif},
  journal={Journal of Atmospheric and Oceanic Technology},
  volume={11},
  number={3},
  pages={661--673},
  year={1994}
}

@article{Lehner2016,
    author = {Manuela Lehner and C. David Whiteman and Sebastian W. Hoch and Erik T. Crosman and Matthew E. Jeglum and Nihanth W. Cherukuru and Ronald Calhoun and Bianca Adler and Norbert Kalthoff and Richard Rotunno and Thomas W. Horst and Steven Semmer and William O. J. Brown and Steven P. Oncley and Roland Vogt and A. Martina Grudzielanek and Jan Cermak and Nils J. Fonteyne and Christian Bernhofer and Andrea Pitacco and Petra Klein},
    title = {The {METCRAX II} Field Experiment: A Study of Downslope Windstorm-Type Flows in {A}rizona’s {M}eteor {C}rater},
    journal = {Bulletin of the American Meteorological Society},
    volume = {97},
    pages = {217--235},
    year = {2016}
    }

@article {MoengSullivan1994,
      author = "Chin-Hoh  Moeng and Peter P.  Sullivan",
      title = "A Comparison of Shear- and Buoyancy-Driven Planetary Boundary Layer Flows",
      journal = "Journal of Atmospheric Sciences",
      year = "1994",
      publisher = "American Meteorological Society",
      address = "Boston MA, USA",
      volume = "51",
      number = "7",
      doi = "10.1175/1520-0469(1994)051<0999:ACOSAB>2.0.CO;2",
      pages=      "999 - 1022",
      url = "https://journals.ametsoc.org/view/journals/atsc/51/7/1520-0469_1994_051_0999_acosab_2_0_co_2.xml"
}

@article{mccoll2016mean,
  title={Mean-velocity profile of smooth channel flow explained by a cospectral budget model with wall-blockage},
  author={McColl, Kaighin A and Katul, Gabriel G and Gentine, Pierre and Entekhabi, Dara},
  journal={Physics of Fluids},
  volume={28},
  number={3},
  year={2016},
 pages ={035107},
  publisher={AIP Publishing}
}

@article{mellor1982development,
  title={Development of a turbulence closure model for geophysical fluid problems},
  author={Mellor, George L and Yamada, Tetsuji},
  journal={Reviews of Geophysics},
  volume={20},
  number={4},
  pages={851--875},
  year={1982},
  publisher={Wiley Online Library}
}

@article{Mosso2024,
author = {Mosso, Samuele and Calaf, Marc and Stiperski, Ivana},
title = {Flux-gradient relations and their dependence on turbulence anisotropy},
journal = {Quarterly Journal of the Royal Meteorological Society},
volume = {150},
number = {763},
pages = {3346-3367},
doi = {https://doi.org/10.1002/qj.4762},
year = {2024}
}

@article{Mosso2025,
author = {Mosso, S. and Lapo, K. and Stiperski, I.},
title = {Revealing the drivers of turbulence anisotropy over complex terrain: an interpretable machine learning approach},
journal = {Boundary Layer Meteorology},
volume = {191},
number={12},
pages = {51},
doi = {10.1007/s10546-025-00946-5},
year = {2025}
}

@article{Mishra2016,
  title = {Sensitivity of flow evolution on turbulence structure},
  author = {Mishra, Aashwin A. and Iaccarino, Gianluca and Duraisamy, Karthik},
  journal = {Phys. Rev. Fluids},
  volume = {1},
  issue = {5},
  pages = {052402},
  numpages = {11},
  year = {2016},
  month = {Sep},
  publisher = {American Physical Society},
  doi = {10.1103/PhysRevFluids.1.052402}
}

@article {MoengWyngaard1989,
      author = "Chin-Hoh  Moeng and John C.  Wyngaard",
      title = "Evaluation of Turbulent Transport and Dissipation Closures in Second-Order Modeling",
      journal = "Journal of Atmospheric Sciences",
      year = "1989",
      publisher = "American Meteorological Society",
      address = "Boston MA, USA",
      volume = "46",
      number = "14",
      doi = "10.1175/1520-0469(1989)046<2311:EOTTAD>2.0.CO;2",
      pages=      "2311 - 2330"
}

@article{Nguyen2013, 
title={Measurements of the budgets of the subgrid-scale stress and temperature flux in a convective atmospheric surface layer}, volume={729}, DOI={10.1017/jfm.2013.302}, journal={Journal of Fluid Mechanics}, author={Nguyen, Khuong X. and Horst, Thomas W. and Oncley, Steven P. and Tong, Chenning}, year={2013}, pages={388–422}}

@article{NguyenTong2015, 
title={Investigation of subgrid-scale physics in the convective atmospheric surface layer using the budgets of the conditional mean subgrid-scale stress and temperature flux}, 
volume={772}, DOI={10.1017/jfm.2015.171}, journal={Journal of Fluid Mechanics}, author={Nguyen, Khuong. and Tong, Chenning}, year={2015}, 
pages={295–329}}

@book{Pope2000,
	Author = {Stephen B. Pope},
	Publisher = {Cambridge University Press},
	Title = {Turbulent Flows},
	Address = {Cambridge, UK},
	Year = {2000}}

@article{Praskovsky1993, title={The sweeping decorrelation hypothesis and energy–inertial scale interaction in high Reynolds number flows},
volume={248}, 
DOI={10.1017/S0022112093000862}, journal={Journal of Fluid Mechanics}, 
author={Praskovsky, A.A. and Gledzer, E.B. and Karyakin, M. Y. and Zhou, A.Y.}, 
year={1993}, 
pages={493–511}
}

@article{qin2025asymptotic,
  title={Asymptotic coefficients of the attached-eddy model derived from an adiabatic atmosphere},
  author={Qin, Yue and Katul, Gabriel G and Liu, Heping and Li, Dan},
  journal={Journal of Fluid Mechanics},
  volume={1011},
  pages={A29},
  year={2025},
  publisher={Cambridge University Press}
}

@book{RotachHoltslag2025,
   title = {Ideal and Real Atmospheric Boundary Layers},
   author = {M. W. Rotach and A.A.M. Holtslag},
   year = {2025},
   publisher = {Academic Press}, 
   address = {London, UK}
}

@article{Rotta1951,
	Author = {Rotta, J.},
	Day = {01},
	Doi = {10.1007/BF01330059},
	Issn = {0044-3328},
	Journal = {Zeitschrift f{\"u}r Physik},
	Month = {Nov},
	Number = {6},
	Pages = {547-572},
	Title = {Statistische {T}heorie nichthomogener {T}urbulenz},
	Url = {https://doi.org/10.1007/BF01330059},
	Volume = {129},
	Year = {1951}
	}

@article{saddoughi1994local,
  title={Local isotropy in turbulent boundary layers at high {R}eynolds number},
  author={Saddoughi, S.G. and Veeravalli, S.V.},
  journal={Journal of Fluid Mechanics},
  volume={268},
  pages={333--372},
  year={1994},
  publisher={Cambridge University Press}
}

@article{so1977model,
  title={On model constants and second order closure for curved shear flows},
  author={So, Ronald MC},
  journal={Applied Scientific Research},
  volume={33},
  number={3},
  pages={353--368},
  year={1977},
  publisher={Springer}
}

@book{Stull88,
	Author = {Roland B. Stull},
	Publisher = {Springer},
	Title = {An Introduction to Boundary Layer Meteorology},
	Year = {1988}}

@article{Stiperski2018,
	Author = {Stiperski, Ivana and Calaf, M},
	Issn = {00359009},
	Journal = {Quarterly Journal of the Royal Meteorological Society},
	Pages = {641--657},
	Title = {Dependence of near-surface similarity scaling on the anisotropy of atmospheric turbulence},
	Volume = {144},
	Year = {2018}}

@article{Stiperski2019,
	Author = {Stiperski, Ivana and Calaf, M and Rotach, M. W.},
	Journal = {Journal of Geophysical Research: Atmosphere},
	Pages = {1428--1448},
	Title = {Scaling, Anisotropy, and Complexity in Near-Surface Atmospheric Turbulence},
	Volume = {124},
	Year = {2019}}

@article{Stiperski2021,
	Author = {Stiperski, Ivana and Katul, Gabriel G. and Calaf, Marc},
	Journal = {Physical Review Letters},
	Title = {Universal return to isotropy of inhomogeneous atmospheric boundary layer turbulence},
	  volume = {126},
  issue = {19},
  pages = {194501},
  doi = {10.1103/PhysRevLett.126.194501},
	Year = {2021}}

@article{stiperski2021convective,
  title={Anisotropy of Unstably Stratified Near-Surface Turbulence},
  author={Stiperski, Ivana and Chamecki, Marcelo and Calaf, Marc},
  journal={Boundary-Layer Meteorology},
  volume={180},
  number={3},
  pages={363--384},
  year={2021},
  publisher={Springer}
}

@article{Stiperski2023,
  title = {Generalizing {M}onin-{O}bukhov Similarity Theory (1954) for Complex Atmospheric Turbulence},
  author = {Stiperski, Ivana and Calaf, Marc},
  journal = {Phys. Rev. Lett.},
  volume = {130},
  issue = {12},
  pages = {124001},
  numpages = {7},
  year = {2023},
  month = {Mar},
  publisher = {American Physical Society},
  doi = {10.1103/PhysRevLett.130.124001}
}

@article{Sfyri2018,
    author = {Eleni Sfyri and Mathias W. Rotach and Ivana Stiperski and Fred C. Bosveld and Lehner Manuela and Friedrich Obleitner},
    journal = {Boundary-Layer Meteorology},
    publisher = {Springer Netherlands},
    title = {Scalar-flux similarity in the layer near the surface layer over mountainous terrain},
    year = {2018},
    volume = {169},
    pages={11--46}}

@article{Salesky2017,
    author = {Salesky, Scott T. and Chamecki, Marcelo and Bou-Zeid, Elie}, 
    journal = {Boundary-Layer Meteorology},
    title = {On the Nature of the Transition Between Roll and Cellular Organization in the Convective Boundary Layer},
    volume = {163},
    pages = {41-68},
    year = {2017}}

@article{Salesky2018, 
title={Buoyancy effects on large-scale motions in convective atmospheric boundary layers: implications for modulation of near-wall processes}, 
volume={856}, 
DOI={10.1017/jfm.2018.711}, journal={Journal of Fluid Mechanics}, author={Salesky, S. T. and Anderson, W.}, 
year={2018}, 
pages={135–168}}

@article{Waterman2025,
Author = {Waterman, T.S. and Stiperski, I. and Chaney, N. and Calaf, M. },
Journal = {Agricultural and Forest Meteorology},
Number = {},
Pages = {},
Title = {Evaluating Anisotropy-based {Monin-Obukhov Similarity Theory} over Canopies and Complex Terrain},
Volume = {},
doi = {10.48550/arXiv.2502.13970},
Year = {2025}
}

@book{Wyngaard2010,
	Author = {John C. Wyngaard},
	Publisher = {Cambridge University Press},
	Title = {Turbulence in the Atmosphere },
	Year = {2010}}

@article{Wyngaardetal1971,
	Author = {Wyngaard, J. and Cot{\'e}, O.},
	Journal = {Journal of the Atmospheric Sciences},
	Number = {2},
	Pages = {190--201},
	Title = {The budgets of turbulent kinetic energy and temperature variance in the atmospheric surface layer},
	Volume = {28},
	Year = {1971}}

@inbook{Wyngaard1973,
    author = "J. C.  Wyngaard",
    title = "Workshop on Micrometeorology",
    publisher = "American Meteorological Society, Boston",
    year = "1973",
    pages = "101-149",
    chapter = "On Surface-Layer Turbulence"
}

@article{Yi2025,
  title = {Nonlinear and buoyancy pressure correlations in stably stratified turbulence},
  volume = {1019},
  ISSN = {1469-7645},
  url = {http://dx.doi.org/10.1017/jfm.2025.10605},
  DOI = {10.1017/jfm.2025.10605},
  journal = {Journal of Fluid Mechanics},
  publisher = {Cambridge University Press (CUP)},
  author = {Yi,  Young Ro and Koseff,  Jeffrey Russell and Bou-Zeid,  Elie},
  year = {2025},
  month = sep 
}

@article{Zeman1981,
author = {Zeman, O},
title = {Progress in the Modeling of Planetary Boundary Layers},
journal = {Annual Review of Fluid Mechanics},
volume = {13},
number = {1},
pages = {253-272},
year = {1981},
doi = {10.1146/annurev.fl.13.010181.001345}
}

@article {Zilitinkevich2021,
      author = "Sergej Zilitinkevich and Evgeny Kadantsev and Irina Repina and Evgeny Mortikov and Andrey Glazunov",
      title = "Order out of Chaos: Shifting Paradigm of Convective Turbulence",
      journal = "Journal of the Atmospheric Sciences",
      year = "2021",
      publisher = "American Meteorological Society",
      address = "Boston MA, USA",
      volume = "78",
      number = "12",
      doi = "10.1175/JAS-D-21-0013.1",
      pages=      "3925 - 3932"
}

\end{document}